\documentstyle[11pt,aasms4]{article}
\def\HI{\hbox{H~$\scriptstyle\rm I\ $}}
\def\HII{\hbox{H~$\scriptstyle\rm II\ $}}
\def\bHII{\hbox{\bf H~$\scriptstyle\bf II\ $}}

\def\HeI{\hbox{He~$\scriptstyle\rm I\ $}}
\def\HeII{\hbox{He~$\scriptstyle\rm II\ $}}
\def\HeIII{\hbox{He~$\scriptstyle\rm III\ $}}
\def\nH{{\rm H}}
\def\nHI{{\rm HI}}

\def\nHeI{{\rm HeI}}

\def\kms{\,{\rm km\,s^{-1}}}
\def\kmsmpc{\,{\rm km\,s^{-1}\,Mpc^{-1}}}

\def\pcc{\,{\rm cm^{-3}}}
\def\kel{\,{\rm K}}
\def\cmm{\,{\rm cm^{-2}}}

\def\uvunits{\,{\rm ergs\,cm^{-2}\,s^{-1}\,Hz^{-1}\,sr^{-1}}}
\def\ccps{\,{\rm cm^{3}\,s^{-1}}}
\def\emunits{\,{\rm ergs\,Mpc^{-3}\,s^{-1}\,Hz^{-1}}}
\def\mpc{\,{\rm Mpc}}
\def\msun{{\rm\,M_\odot}}
\def\CBR{{\rm CBR}}

\def\Lya{Ly$\alpha\ $}

\def\etal{{et al.\ }}
\def\spose#1{\hbox to 0pt{#1\hss}}
\def\lsim{\mathrel{\spose{\lower 3pt\hbox{$\mathchar"218$}}
     \raise 2.0pt\hbox{$\mathchar"13C$}}}
\def\gsim{\mathrel{\spose{\lower 3pt\hbox{$\mathchar"218$}}
     \raise 2.0pt\hbox{$\mathchar"13E$}}}
 
\begin{document}
\title{21-CM TOMOGRAPHY OF THE INTERGALACTIC MEDIUM AT HIGH REDSHIFT}

\author{Piero Madau}

\affil{Space Telescope Science Institute, 3700 San Martin Drive,
Baltimore MD 21218;\ madau@stsci.edu}

\author{Avery Meiksin\altaffilmark{1}}

\affil{Dept. of Astronomy and Astrophysics, University of Chicago,
5640 S. Ellis Ave., Chicago IL 60637;\ meiksin@oddjob.uchicago.edu} 

\altaffiltext{1}{Edwin Hubble Research Scientist}

\and

\author{Martin J. Rees}

\affil{Institute of Astronomy, Madingley Road, Cambridge CB3 0HA}

\begin{abstract}
We investigate the 21-cm signature that may arise from the intergalactic
medium (IGM) prior to the epoch of full reionization ($z>5$). In scenarios
in which the IGM is reionized by discrete sources of photoionizing radiation,
the neutral gas which has not yet been engulfed by an \HII region may easily
be preheated to temperatures well above that of the cosmic background radiation
(CBR), rendering the IGM invisible in absorption against the CBR. We identify
three possible preheating mechanisms:\ (1)\ photoelectric heating by soft
X-rays from QSOs, (2)\ photoelectric heating by soft X-rays from early galactic
halos, and (3)\ resonant scattering of the continuum UV radiation from an early
generation of stars. We find that bright quasars with only a small fraction of
the observed comoving density at $z\sim 4$ will suffice to preheat the entire
universe at $z\gsim 6$. We also show that, in a Cold Dark Matter dominated
cosmology, the thermal bremsstrahlung radiation associated with collapsing
galactic mass halos ($10^{10}-10^{11}\msun$), may warm the IGM to
$\sim 100\kel$ by $z\sim7$. Alternatively, the equivalent of $\sim 10\%$ of the
star formation rate density in the local universe, whether in isolated
pregalactic stars, dwarf, or normal galaxies, would be capable of heating the
entire IGM to a temperature above that of the CBR by \Lya scattering in a small
fraction of the Hubble time at $z\sim 6$. 

In the presence of a sufficiently strong ambient flux of \Lya photons, the
hyperfine transition in the warmed \HI will be excited. A beam differencing
experiment would detect a patchwork of {\it emission}, both in frequency and
in angle across the sky. This patchwork could serve as a valuable tool for
understanding the epoch, nature, and sources of the reionization of the
universe, and their implications for cosmology. We demonstrate that isolated
QSOs will produce detectable signals at meter wavelengths within their
``spheres of influence'' over which they warm the IGM. As a result of the
redshifted 21-cm radiation emitted by warm \HI bubbles, the spectrum of the
radio extragalactic background will display frequency structure with velocity
widths up to $10,000\kms$. Broad beam observations would reveal corresponding
angular fluctuations in the sky intensity with $\delta T/T\lsim 10^{-3}$ on
scales $\theta\sim 1^\circ$. This scale is set either by the ``thermalization
distance'' from a QSO within which \Lya pumping determines the spin temperature
of the IGM or by the quasar lifetime. 

Radio measurements near 235 and $150\,$MHz, as will be possible in the near
future using the {\it Giant Metrewave Radio Telescope}, may provide the first
detection of a neutral IGM at $5\lsim z\lsim 10$. A next generation facility
like the {\it Square Kilometer Array Interferometer} could effectively open
much of the universe to a direct study of the reheating epoch, and possibly
probe the transition from a neutral universe to one that is fully ionized. 
 
\end{abstract}


\section{Introduction}

At epochs corresponding to $z\sim 1000$, the intergalactic medium (IGM)
is expected to recombine and remain neutral until sources of radiation develop
that are capable of reionizing it. The application of the Gunn-Peterson (1965)
constraint on the amount of cosmologically distributed neutral hydrogen to QSO
absorption spectra requires the universe  to have been highly ionized by $z\sim
5$ (Schneider, Schmidt, \& Gunn 1991). It thus appears that substantial sources
of ionization were present at these or earlier epochs, perhaps QSOs and young
galaxies, or some as yet undiscovered class of objects such as Pop III stars or
decaying particles. Establishing the epoch of reionization is crucial for
determining the impact of reionization on several key cosmological issues, from
the role reionization plays in allowing collapsed objects to cool and
make stars, to determining the small-scale structure in the temperature
fluctuations of the cosmic background radiation (CBR) (e.g., Sugiyama, Silk,
\& Vittorio 1993; Bond 1995; Hu \& White 1996). Conversely, probing the epoch
of reionization may provide a means of detecting the onset of the first
generation of stars.

Very little is known about the nature of the first bound objects and
of the thermal state of the universe at early epochs. In particular,
the history of the transition from a neutral IGM to an almost fully
ionized one can reveal the character of the ionizing sources. If these
are uniformly distributed, like an abundant population of pregalactic
stars or decaying dark matter, the ionization and thermal state of the
IGM will be the same everywhere at any given epoch, with the neutral
fraction decreasing gradually with cosmic time.  In a CDM dominated
cosmology, this may be the case since bound objects sufficiently
massive ($\sim 10^6 \msun$) to make stars form at high redshift.  If a
sufficient fraction of their ionizing photons are able to escape, the
stars in these objects would reionize the universe by $z\gsim 20$
(Couchman \& Rees 1986). In inhomogeneous
reionization scenarios instead, widely separated but very luminous
sources of photoionizing radiation such as QSOs, present at the time
the IGM is largely neutral, will generate expanding \HII regions whose
size is only limited by the individual source lifetime. The universe
will be divided into an ionized phase whose filling factor $q(z)$
increases with time, and an ever shrinking neutral phase (Arons \&
Wingert 1972; Meiksin \& Madau 1993). If the ionizing sources are
randomly distributed, the \HII regions will be spatially isolated for
$q\ll 1$. The IGM will become completely reionized at the breakthrough
epoch, when $q=1$ and the IGM is fully transparent at the Lyman
continuum.  Current QSO counts suggest that the breakthrough epoch may
occur as recently as $z\gsim 5$ (Meiksin \& Madau 1993).

It is therefore of great interest, in testing rival reionization
scenarios, to seek any evidence of non-uniformity in the ionization
structure of the IGM by studying the thermal and ionization history of
the intergalactic gas surrounding isolated sources of radiation (see,
e.g., Miralda-Escud\`e \& Rees 1994). In this paper we shall discuss
how radio-astronomical measurements of spectral and spatial structures
in the 21-cm line emission and absorption from \HI at high redshifts
may provide a valuable tool for probing the epoch of the likely heat
input at $5\lsim z\lsim 10$ and its nature, e.g., whether it was
limited to small isolated regions or was relatively diffuse. We will
confine our analysis to epochs prior to complete reionization, and
address the physical state of the gas which has not yet been engulfed
by an ionized region.

For intergalactic \HI to be observable in 21-cm line emission or absorption
against the cosmic background radiation, the spin temperature must differ from
the CBR temperature. As the former will in general be a weighted mean between
the matter and CBR temperatures, an efficient mechanism for coupling the
populations of the hyperfine levels to the kinetic state of the intergalactic
gas is needed to unlock the spin state from the CBR.  At the low particle
densities characteristic of the IGM, the dominant coupling mechanism is
level-mixing via \Lya photon scattering -- the ``Wouthuysen-Field'' effect. 
We present in this paper a new heating mechanism, arising from atomic
recoil in response to the scattering of resonant photons. We demonstrate that
because of resonant scattering heating, \Lya coupling makes the detection of
the IGM in absorption against the CBR quite unlikely, since the same \Lya
photons that pump the hyperfine levels heat the atoms as well. This is not
entirely surprising, as the energetic demand for raising the kinetic
temperature above that of the CBR is relatively small, $\sim
10^{-7}\,$eV$\,$cm$^{-3}$ at $z\sim 7$, hence even a relatively inefficient 
heating mechanism like \Lya resonant scattering may become important.

We find that the heating requirements are quite generally met in a
variety of scenarios in which radiation sources turn on at high
redshifts. We show that, in models based on QSO photoionization, and
well before the \HII region network has fully percolated, the (mostly)
neutral IGM between the \HII bubbles will be photoelectrically heated
to temperatures above a few hundred degrees by soft X-ray photons from
the QSOs. In Cold Dark Matter (CDM) models a similar mechanism may
efficiently warm the IGM at $z\lsim 7$, since galactic mass halos with
virial temperatures $\sim 10^6\,$K are significant sources of thermal
bremsstrahulung radiation. Alternatively, if protogalaxies at these
early epochs are forming stars at a rate which is only $\sim 10\%$ of
the global present-day star formation rate, heating by background
Ly$\alpha$ ``continuum" photons could drive the kinetic temperature of
the still neutral material well above the radiation temperature of the
microwave background.  As a consequence, we conclude that {\it most of
the neutral IGM will be available for detection at 21-cm only in
emission.}

Two scenarios serve to illustrate the observability at meter
wavelengths of the intergalactic medium during reionization. An
isolated QSO turning on in a neutral medium will create an \ion{H}{2}
bubble surrounded by an X-ray warmed neutral zone. Sufficiently near
the QSO, the spin state of the neutral IGM will be coupled to its
kinetic state by redshifted continuum \Lya photons from the QSO. This
zone will then emit 21-cm radiation that would be detectable against
the CBR as a differential intensity, in angle or frequency. A stronger
signal may be achieved in the presence of a \Lya continuum background
sufficiently strong to couple the spin temperature of the cold and
neutral IGM outside the QSO ``sphere of influence'' to the kinetic
temperature of the IGM. The larger signal results from the enhanced
contrast between the 21-cm absorbing medium far from the QSO, and the
21-cm emitting gas in the X-ray warmed neutral zone surrounding the
QSO \HII region. This phase, however, will be short-lived: through
atomic recoil the \Lya background photons will soon heat the IGM far
from the QSO to a temperature above that of the CBR, and absorption
will cease.

The plan of the paper is as follows. In \S~2 we discuss sources of
radiation at high redshift. In \S~3 we review the mechanisms that
determine the level populations of the hyperfine state of neutral
hydrogen. In \S~4 we treat the preheating of the IGM prior to
complete reionization, and show that a 21-cm absorbing IGM can exist only
as a transitory state. In \S~5 we argue that the 21-cm line
signature from a neutral IGM, redshifted to meter wavelengths, will
appear as a small feature in the radio extragalactic background on
scales of $\lsim 1^\circ$, and may be detectable with the {\it Giant Metrewave
Radio Telescope}.  Finally, in \S~6, we summarize our conclusions. 

\section{Sources of Radiation at High Redshift}

In the following, we assume that the universe is completely reionized by
discrete sources at $z\gsim5$, and that the IGM at such early epochs is uniform
on large scales. Unless otherwise stated, we shall adopt a flat cosmology with
$q_0=0.5$ and $H_0=50\,h_{50}\kmsmpc$. 

\subsection{QSO \bHII Regions}

When an isolated point source of ionizing radiation turns on, the
ionized volume initially grows in size at a rate fixed by the emission of
UV photons, and an ionization front separating the ionized and neutral
regions propagates into the neutral gas.  The evolution of an expanding 
cosmological I-front in a uniform IGM is governed by the equation 
\begin{equation}
4\pi r_I^2n_{\nH}\left({{dr_I}\over{dt}}-Hr_I\right) =
S - {4\over3}\pi r_I^3 (1+2\chi) n_{\nH}^2\alpha_B, \label{eq:dridt}
\end{equation}
(Shapiro 1986), where $r_I$ is the proper radius of the I-front, 
$n_{\nH}$ is the hydrogen density of the IGM, $H$ is the Hubble constant, 
$S$ is the number of ionizing photons emitted by the central source per unit 
time, $\alpha_B\simeq2.6 \times 10^{-13}T_4^{-0.845}$~cm$^3$~s$^{-1}$ is the
recombination coefficient to the excited states of hydrogen, and $T_4$ is the 
gas temperature in units of $10^4\,$K. A helium
to hydrogen cosmic abundance ratio $\chi=1/12$ is adopted, with both H and He
fully ionized inside the \HII region.  The right hand side of
equation~(\ref{eq:dridt}) is a measure of the net flux of ionizing photons
reaching the I-front
after subtracting recombinations, and vanishes at the Str\"omgren radius,
$r_{\rm St}=\{3S/[4\pi n^2_\nH\alpha_B(1+\chi)]\}^{1/3}$. Across the
I-front the degree of ionization changes sharply on a distance of the order
of the mean free path of an ionizing photon. 
Denoting by $\Omega_{\rm IGM}$ the baryonic mass density parameter of 
the intergalactic medium, the I-front will expand at the speed of light 
\begin{equation}
r_I<\left({S\over 4\pi n_\nH c}\right)^{1/2}\approx (2.85\,{\rm Mpc}) 
\left({S\over 10^{57}\,{\rm s}^{-1}}\right)^{1/2}
\left({\Omega_{\rm IGM}h_{50}^2\over 0.05}\right)^{-1/2}
\left({1+z\over 7}\right)^{-3/2}, \label{eq:rcrit}
\end{equation}
to subsequently slow down because of geometrical dilution. In an expanding IGM,
the ionization front generated by a QSO, with $S\sim10^{57} \rm s^{-1}$, or
a star-forming galaxy, with $S\sim10^{53} \rm s^{-1}$, will fail to grow to
even half of its Str\"omgren radius (Shapiro 1986; Madau \& Meiksin 1991). In
this case radiative recombinations within the \HII region may be neglected on
the right-hand side of equation~(\ref{eq:dridt}) (proportional to 
$r_{\rm St}^3-r_I^3$), to obtain at a short time after the turn-on epoch $z$
(corresponding to the redshift interval $\Delta z\ll 1+z$),
\begin{equation}
r_I\approx (10.2\,{\rm Mpc}) \left({S\over 10^{57}\,{\rm s}^{-1}}\right)
^{1/3}h_{50}^{-1/3}\left({\Omega_{\rm IGM}h_{50}^2\over 0.05}\right)^{-1/3}
\left({1+z\over 7}\right)^{-11/6}\Delta z^{1/3}, \label{eq:rieq}
\end{equation}
and
\begin{equation}
v_I\approx (2.2\times 10^4{\kms})\left({S\over 10^{57}\,{\rm s}^{-1}}\right)
^{1/3}h_{50}^{2/3}\left({\Omega_{\rm IGM}h_{50}^2\over 0.05}\right)^{-1/3}
\left({1+z\over 7}\right)^{2/3}\Delta z^{-2/3}. \label{eq:vieq}
\end{equation}
For comparison, the radius of the light front around each source is
\begin{equation}
r_c\equiv c\Delta t\approx (46.3\,{\rm Mpc}) h_{50}^{-1}
\left({1+z\over 7}\right)^{-5/2}\Delta z.
\end{equation}

\subsection{Virialized Halos}

In CDM dominated cosmologies, galactic mass halos will begin to
collapse at high redshift ($z>5$). The hot gas in these halos will
produce a soft X-ray background radiation field via thermal
bremsstrahlung.  We estimate the number of halos using the
Press-Schechter formalism (Press \& Schechter 1974), which should
suffice for an order of magnitude estimate for systems on these
scales.

The (proper) volume emissivity per unit frequency due to thermal bremsstrahlung
radiation from isothermal, optically thin halos is
\begin{eqnarray}
\epsilon_{\rm halo}(\nu) = (3.2\times10^{36}\emunits)
\int_0^\infty & dr_0 N(r_0,z)(1+z)^3 n_{\rm halo}^2(r_0) {{4\pi}\over3}r_v^3
T(r_0)^{-1/2} \nonumber \\ & \times \exp[-h\nu/kT(r_0)], \label{eq:eb_halo}
\end{eqnarray}
where $N(r_0,z)$ is the number of halos per unit comoving volume per initial
comoving size $r_0$ at redshift $z$, $n_{\rm halo}(r_0)$ is the internal
hydrogen density of a halo, and $T(r_0)$ is its temperature. The hydrogen is
assumed to be shock heated to the virial temperature and collisionally ionized.
In spherical top-hat collapse, the temperature $T$ and radius $r_v$ of a
virialized halo are related to $r_0$ by $kT(r_0)=1.4\mu m_{\rm H}(1+z_c)H_0^2
r_0^2$ and $r_0=5.6(1+z_c)r_v$, where $z_c$ is the collapse redshift of the
perturbation, and $\mu\approx0.6$ is the mean molecular weight of 
a fully ionized gas of cosmic abundances. The internal density of 
the cloud exceeds the external density at
the projected time of collapse by a factor of 178. Normalizing to a baryon
density of $\Omega_{\rm IGM}h_{50}^2=0.05$ gives $n_{\rm
halo}=1.8\times10^{-5}(1+z_c)^3\pcc$. According to the Press-Schechter
formalism,
\begin{equation}
N(r_0,z)dr_0 =-3{1.686(1+z)\over (2\pi)^{3/2}r_0^4\Delta(r_0)}
{d\log\Delta(r_0)\over d\log(r_0)}\exp\left[-{1.686^2(1+z)^2
\over 2\Delta(r_0)^2}\right]dr_0.
\end{equation}
Here, $\Delta(r_0)$ is the rms mass fluctuation
in a sphere of radius $r_0$. We will adopt a pure CDM spectrum for a $q_0=0.5$,
$H_0=50\kmsmpc$ universe (Bardeen \etal 1986). This model 
appears to provide an adequate representation of the power spectrum on small
scales when appropriately renormalized, even though a higher normalization
is needed for an optimal match to the temperature fluctuations measured
by COBE on large scales (G\' orski \etal 1995). Normalizing to the abundance of
clusters, $\Delta(16\mpc)\approx 0.6$ (White \etal 1993), we find that for $0.2
<r_0<1\mpc$, $\Delta(r_0)\approx 3.6r_0^{-3/8}$, where $r_0$ is measured in
megaparsecs. 

The total emissivity above a minimum photon energy $h\nu_{\rm min}$ is 
$\epsilon_{\rm halo}=[kT(r_0)/h]\epsilon_{\rm halo}(\nu_{\rm min})$, after
integrating equation~(\ref{eq:eb_halo}) over frequency. While the emissivity
from an
individual halo increases for increasing temperature (for $kT<h\nu_{\rm min}$),
the number of halos decreases rapidly. At any given redshift, the peak
contribution to the integral in equation~(\ref{eq:eb_halo}) arises therefore
from some intermediate size halo. We may thus approximate this integral 
by performing a second order Taylor expansion of the exponent about the 
halo size. Most of the bremsstrahlung radiation is found to be emitted 
by halos with  an initial comoving radius
\begin{equation}
\hat r_0\approx (1\mpc) \left({{h\nu_{\rm min}}\over{20\,{\rm ryd}}}
\right)^{4/11}\left({{1+z}\over7}\right)^{-12/11}. \label{eq:rpeak}
\end{equation}
The total emissivity is given by
\begin{eqnarray}
\epsilon_{\rm halo}\approx(2.6\times10^{43}{\rm\, ergs\, Mpc^{-3}\,
s^{-1}\,}) &  {\hat r}_0^{3/8}\left({{1+z}\over7
}\right)^7 \left[{{kT(\hat r_0)}\over{20\,{\rm ryd}}}\right] \left({{h\nu_{\rm
min}}\over{20\,{\rm ryd}}}\right)^{-1/2} \nonumber \\ & \times
\exp\left[-5.3\left({{1+z}\over7}\right)^2\hat r_0^{3/4}-
1.8\left({{1+z}\over7}\right)^{-1}\left({{h\nu_{\rm min}} \over{20\,{\rm
ryd}}}\right)\hat r_0^{-2}\right]. \label{eq:ehalo} 
\end{eqnarray}
We shall use this expression in \S~4.2.3 below to estimate the virialized halo
contribution to the heating rate of the IGM.

\subsection{Early Stellar Populations}

An early generation of stars may provide a background radiation field
of \Lya photons that arises from the collective redshifted continua of
the stars. The background specific intensity associated with a constant
comoving emissivity from the present epoch to redshift $z_{\rm max}$, 
as seen by an observer at redshift $z$, is given by
\begin{equation}
J_\alpha(z)={1\over 4\pi} \epsilon(\nu_\alpha,0)(1+z)^{3+\alpha_S}
\int_{z}^{z_{\rm max}} {d\ell\over dz'}(1+z')^{-\alpha_S} dz', \label{eq:eqJ}
\end{equation}
where $\epsilon(\nu_\alpha,0)$ is the \Lya volume emissivity due to stars
in galaxies at the present epoch, $d\ell/dz'$ is the proper differential
line element in a Friedmann cosmology, and we have assumed 
a power law energy spectrum. Here, $1+z_{\rm max}\approx 4(1+z)/3$, as photons
emitted at higher redshifts will have wavelength below $912\,$\AA~ at the
source, and will be strongly attenuated by Lyman-continuum intergalactic
absorption. Weakly dependent on $\alpha_S$, equation~(\ref{eq:eqJ}) may be
rewritten as  
\begin{equation}
J_{\alpha}(z)\approx (10^{-20.7}\uvunits)~h_{50}^{-1}
\left[{\epsilon(\nu_\alpha,0)\over 10^{25}\,{\rm ergs\, Mpc^{-3}\,
s^{-1}\, Hz^{-1}}}\right]\left({1+z \over 7}\right)^{3/2}. \label{eq:Jast}
\end{equation}
Using the results of the H$\alpha$ survey of Gallego \etal (1995), 
we estimate the continuum emissivity from present day galaxies at the
frequency of \Lya to be, to within a factor of 2,
$\epsilon(\nu_\alpha,0)\approx 4\times10^{25}\,$ergs Mpc$^{-3}$ s$^{-1}$
Hz$^{-1}$. Only a fraction of this emissivity would be available at high
redshift from dwarf galaxies or the progenitors of today's bright galaxies. The
\Lya continuum emissivity at early epochs, however, may be even larger if
pregalactic Pop III stars are present at these times. 

The IGM will still be mostly
neutral if the background intensity at the Lyman edge, $J_L$, satisfies
\begin{equation}
J_L<{hc\over 4\pi}n_\nH(0)(1+z)^3\approx (10^{-21.2}\uvunits)
\left({\Omega_{\rm IGM}h_{50}^2\over 0.05}\right)
\left({1+z\over 7}\right)^3, \label{eq:JL}
\end{equation}
i.e., if less than one ionizing photon per hydrogen atom has reached the IGM by
that epoch. Because the integrated UV spectrum of a stellar population is
characterized by a strong Lyman-continuum break (about a factor of 4), and,
more importantly, only a small fraction of ionizing photons might escape from
collapsed star-forming regions into intergalactic space (cf. Madau \& Shull
1996), one expects $J_L\ll J_\alpha$. Hence, based on a comparison of
equations (\ref{eq:Jast}) and (\ref{eq:JL}), it is plausible to assume a
largely neutral medium even in the presence of a significant \Lya background
from an early generation of stars, $J_{\alpha,-21}\approx 1$. 

\section{Spin Temperature of the IGM}

The amount of emission or absorption from a neutral intergalactic medium is
determined by the spin state of the atoms. The spin temperature may
be coupled to the matter both through the scattering of \Lya photons -- the
Wouthuysen-Field effect (Wouthuysen 1952; Field 1958) -- and atomic collisions.
At the low particle densities characteristic of intergalactic gas, and in the
presence of radio-quiet radiation sources, the dominant coupling mechanism is
\Lya scattering. 

\subsection{\Lya Radiation Color Temperature}

The Wouthuysen-Field effect mixes the hyperfine levels of neutral hydrogen via
the intermediate step of transitions to the $2p$ state. According to this
mechanism, an atom initially in the $n=1$ $_0S_{1/2}$ singlet state may absorb
a \Lya photon that will put it in the $n=2$ $_1P_{1/2}$ or $_1P_{3/2}$ state,
allowing it to return to the triplet $n=1$ $_1S_{1/2}$ state by spontaneous
decay. Similarly, a triplet atom may absorb a slightly lower frequency photon
to reach the same upper state, followed by decay to the singlet state. The
efficiency of the effect is determined by the frequency dependence of the
background radiation field $J_\nu$ near \Lya. In particular, the excitation
rates of the hyperfine levels depend on the relative intensities of the blue
and the red wings of \Lya. To observe the 21-cm line in absorption or emission,
the excitation and de-excitation rates of the hyperfine levels must be
comparable to or exceed those due to 21-cm continuum background radiation. 

The \Lya line absorption cross section is
\begin{equation} 
\sigma_\nu=\sigma_\alpha\phi(\nu)={\pi e^2 \over m_e
c}f_{\alpha}\phi(\nu)={3\over 8 \pi}\lambda_\alpha^2 A_{\alpha}\phi(\nu)
\label{eq:cross}
\end{equation}
(neglecting the very small correction due to stimulated emission, of order
$\lambda_\alpha^3J_\alpha/2hc\ll1$), where $A_{\alpha}$ and $f_{\alpha}$ are
the spontaneous Einstein coefficient and upward oscillator strength for the
transition, $\phi(\nu)$ is the line profile function (with normalization
$\int\phi(\nu) d\nu=1$), and all other symbols have their usual meanings. 
The absorption rate of \Lya photons is proportional to $B_{jk}\int d\nu
\phi(\nu)J_\nu$, where $B_{jk}$ is the Einstein stimulated absorption
coefficient from level $j$ to level $k$, where we now mean to include the
hyperfine structure of the atom. [The coefficients are related to the total
spontaneous decay rates $A_{kj}$ between the various hyperfine levels by
$B_{jk}=(\lambda_{jk}^3/ 8\pi h)(g_k/g_j)A_{kj}$.] When the intensity varies
smoothly across the \HI absorption profile for \Lya, that is, over at least one
Doppler width $\Delta\nu_{\rm D}=(b/c)\nu_\alpha$,
where $b=(2kT_K/m_{\rm H})^{1/2}$ and $\nu_\alpha$ is the line center
frequency,
it is useful to express the effect of the mixing mechanism in
terms of the color temperature $T_\alpha$ of the radiation field, which we
define by 
\begin{equation}
{1\over {kT_\alpha}}=-{{\partial \log{\cal N}_\nu}\over{\partial h\nu}},
\end{equation}
evaluated at $\nu=\nu_\alpha$, where ${\cal N}(\nu)$ is the photon occupation
number at frequency $\nu$, ${\cal N}(\nu)=c^2 J_\nu/(2h\nu^3)$. If the
background continuum spectrum near \Lya is a power law,
$J_\nu\propto\nu^{-\alpha_S}$, then 
$T_\alpha=(h\nu_\alpha/k)/(3+\alpha_S)\approx 3\times 10^4\,{\rm K}
[4/(3+\alpha_S)]$.
It has been shown by Field (1959b), however, that, because of the large 
cross section for resonant scattering, the shape of the 
radiation spectrum near \Lya will be determined in all the relevant
physical situations by the kinetic temperature of the scattering material.
This tendency for thermalization is caused by the recoil of the atom, 
which shifts the profile to the red. As stimulated emission is negligible, the
radiation behaves like a classical relativistic Boltzmann gas, and the specific
number density of photons, $n_\nu=4\pi J_\nu/ch\nu$, assumes the quasi-LTE form
$n_\nu={\rm const}\times \exp [-h(\nu-\nu_\alpha)/kT_K]$, where $T_K$ is the
kinetic temperature of the gas, to which the color temperature relaxes. 

The influence of cosmological expansion on the \Lya profile has been
discussed by Field (1959a). The effect of expansion is determined by
two timescales, the mean-free scattering time for a \Lya photon,
$b/(cn_\nHI\sigma_\alpha\lambda_\alpha)$, where $n_{\rm HI}$ is the
density of neutral hydrogen, and the time it takes the Hubble
expansion to redshift a photon through the resonance, $b/(Hc)$.  Their
ratio, also known within the framework of line radiation transfer
theory as the Sobolev parameter (and equal in this context to the
inverse of the \Lya optical depth), is
\begin{equation}
\gamma\equiv{H\over \lambda_\alpha\sigma_\alpha n_\nHI}
\approx6\times10^{-6}h_{50}\left({1+z\over 7}\right)^{-3/2}\left({\Omega_{\rm
IGM}h_{50}^2 \over 0.05}\right)^{-1}.
\end{equation}
As this is much
smaller than unity, the \Lya part of the radiation background spectrum
establishes close thermal contact with the expanding IGM, and a
quasi-static treatment of the cosmological line transfer is an
adequate approximation (see, e.g., Rybicki \& Dell' Antonio 1994). The
profile of \Lya radiation in an expanding cloud has been calculated in
detail by Deguchi \& Watson (1985). These authors consider the case in
which \Lya photons are created by recombinations inside the cloud.
They show that $T_\alpha\rightarrow T_K$ for $\gamma\lsim 10^{-5}$,
similar to what is found in the absence of velocity gradients (Field
1959b).  The color temperature approaches the gas kinetic temperature
even more quickly if the cloud is irradiated by continuum radiation in
the vicinity of \Lya, as this spectrum is already flat in frequency,
and only a small number of scatterings is needed for the radiation
field to achieve the proper slope (Deguchi \& Watson 1985). In the
next section we shall see how, in the presence of a strong ambient
flux of \Lya photons, the color temperature governs the hyperfine
populations of atomic hydrogen.

\subsection{\Lya Photon Pumping}

In the absence of collisions, the spin temperature, $T_S$, of the 
hyperfine levels of intergalactic hydrogen (as defined by the Boltzmann 
equation) is determined solely by the scattering of ambient \Lya photons and 
by the microwave background radiation field at 21-cm (Field 1958, 1959a).
\footnote{This is not strictly true in the presence of bright
continuum sources of 21-cm radiation, such as radio-loud quasars (see
\S~3.5.).}~In a steady state we have
\begin{equation}
T_S={T_\CBR+y_\alpha T_\alpha\over 1+y_\alpha}, \label{eq:tspin}
\end{equation}
where $T_\CBR=2.73(1+z)\,$K is the temperature of the cosmic background
radiation (Mather \etal 1994), and
\begin{equation}
y_\alpha\equiv{P_{10}\over A_{10}}{T_*\over T_\alpha} \label{eq:ya}
\end{equation}
is the normalized probability, or \Lya pumping ``efficiency''. Here, 
$T_*\equiv h\nu_{10}/k=0.07\,$K, $A_{10}=2.9\times 10^{-15}\,$s$^{-1}$ is the 
spontaneous decay rate of the hyperfine transition of atomic hydrogen, $P_{10}$
is the indirect de-excitation rate of the triplet via absorption of a \Lya
photon to the $n=2$ level, and we have assumed $T_S\gg T_*$. In the absence of
\Lya pumping the spin temperature goes to equilibrium with the 21-cm background 
radiation field on a timescale $T_*/(T_\CBR A_{10})\approx 5\times
10^4\,$yr, and neutral intergalactic hydrogen cannot be seen in either 21-cm
emission or absorption. If $y_\alpha$ is large, $T_S\rightarrow
T_\alpha\approx T_K$, signifying equilibrium with the matter. The $1\rightarrow
0$ transition rate via \Lya can be replaced by the total rate $P_\alpha$,
\begin{equation}
P_\alpha=\int d\Omega \int {I_\nu\over h\nu}\sigma_\nu d\nu,
\label{eq:qsopalpha}
\end{equation}
at which \Lya photons are scattered by an H atom in the gas,
$P_{10}=4P_\alpha/27$. For a point source, $I_\nu=f_\nu\delta(\Omega)$, where
$f_\nu$ is the specific flux from the source. For an isotropic radiation field,
$I_\nu=J_\nu$. In the limit $T_\alpha\gg T_\CBR$, the fractional deviation in a
steady state of the spin temperature from the temperature of the microwave
background is 
\begin{equation}
{{T_S-T_\CBR}\over T_S}\approx \left[1+{T_\CBR\over y_\alpha
T_\alpha}\right]^{-1}. \label{eq:ftspin}
\end{equation}
There exists then a critical value of $P_\alpha$ which, if much exceeded, would
drive $T_S\rightarrow T_\alpha$. This ``thermalization rate'' is found to be
\begin{equation}
P_{\rm th}\equiv {27A_{10}T_\CBR\over 4T_*}\approx 
(5.3\times 10^{-12} ~{\rm s}^{-1})~ \left({1+z\over 7}\right).
\label{eq:ptherm}
\end{equation}
When $P_\alpha=P_{\rm th}$, we have $T_S=2T_\CBR/ (1 + T_\CBR/ T_\alpha)$.
It should be emphasized that, while $T_\alpha$ depends on the spectral gradient
across line center, $y_\alpha T_\alpha$, and hence $P_{\rm th}$, are only 
functions of the magnitude of $J_\alpha$, independent of the uncertainties 
regarding the actual profile of the \Lya background photon spectrum. 

\subsection{Level Mixing by \Lya Continuum Photons}

There are four possible sources of background \Lya photons propagating into
the (mostly) neutral material which has not yet been engulfed by an \HII
region: a) UV, non-ionizing continuum emitted by the radiating sources
and then redshifted by the Hubble expansion to the \Lya wavelength in
the absorbing medium rest-frame; b) \Lya photons generated by radiative
recombinations and escaping from the \HII regions; c) \Lya produced
locally in the warm neutral material surrounding the ionized gas
by radiative recombinations; and d) by collisional excitations.
Of these, only the first, continuum photons redshifted into local \Lya, is
a significant source of coupling, and we treat it next. We show that the
remaining three are generally negligible in Appendix A.

In the limit where the scattering occurs close to
the emission redshift, we find for a point source
\begin{equation}
P_\alpha\approx (2.3\times 10^{-10}~{\rm s}^{-1})~r_{\rm Mpc}^{-2}
L_{\alpha,47}, \label{eq:plpalpha}
\end{equation}
where $L_{\alpha,47}$ is the source \Lya luminosity per octave in photon energy
in units of $10^{47}\,$ ergs s$^{-1}$, and $r_{\rm Mpc}$ is the (physical)
distance in Mpc between emission and scattering. Thus, within several Mpc from
a bright, radio-quiet QSO, $P_\alpha>P_{\rm th}$ and the spin temperature of
neutral hydrogen will be determined by the \Lya continuum photons emitted from
the quasar itself.

Alternatively, we may consider the effect of an integrated intensity from a
distribution of sources. The \Lya scattering rate is then
\begin{equation}
P_\alpha\approx (8.5\times 10^{-12}~{\rm s}^{-1})~ J_{\alpha,-21},
\label{eq:pj}
\end{equation}
where $J_{\alpha,-21}$ is the intensity of the background diffuse field
at \Lya in units of $10^{-21}\uvunits$. For $J_{\alpha,-21}\ll 1$, the spin
system will go into equilibrium with the microwave background.
Equation~(\ref{eq:pj}) is also approximately valid in
the case of recombination \Lya photons, as radiative transfer calculations
predict $dJ_\nu/d\nu=0$ at the Doppler core of the transition (Urbaniak \&
Wolfe 1981). Spectral gradients induced by atomic recoil only result in higher
order corrections for $P_\alpha$. The rate is therefore dependent only on
$J_\alpha$. 

The \Lya pumping mechanism poses a difficulty for detecting the IGM in
absorption against the CBR, since, as we will show below, the same \Lya photons
that mix the hyperfine levels of hydrogen heat the atoms as well. Before we
discuss this problem, we briefly address the influence of particle
collisions and direct radio continuum on the spin temperature.

\subsection{Hyperfine Excitation via Particle Collisions}

We demonstrate in this section that, except at very high redshifts, or in 
overdense regions, collisions are not effective in coupling the spin 
temperature to the kinetic temperature. One can show that, in the general 
case, the spin temperature may be written in the form (Field 1958) 
\begin{equation}
T_S={T_\CBR+y_\alpha T_\alpha+y_c T_K\over 1+y_\alpha+y_c}, \label{eq:tspinc}
\end{equation}
where
\begin{equation}
y_c\equiv{C_{10}\over A_{10}}{T_*\over T_K},
\end{equation}
and $C_{10}$ is the rate of collisional de-excitation of the triplet level
(cf. eqs. [\ref{eq:tspin}] and [\ref{eq:ya}]). Emission occurs for 
\begin{equation}
{y_\alpha T_\alpha + y_c T_K \over y_\alpha + y_c}\approx T_K>T_\CBR,
\end{equation}
and absorption otherwise. Unless the electron density is larger than a few
percent of the hydrogen density, the collisional de-excitation rate is
dominated by collisions with other hydrogen atoms through spin-exchange
(Purcell \& Field 1956). Allison \& Dalgarno (1969) give a maximum
de-excitation rate of $C_{10}/n_{\rm H}=3.3\times10^{-10}\ccps$ for
$T_K<1000\kel$, with a value a factor 2.6 lower at $T_K=100\kel$. To couple the
spin temperature to the kinetic temperature requires a collision rate
$C_{10}>A_{10}T_\CBR/T_*$, or a density $n_{\rm H}>3\times10^{-4}
(1+z)\pcc$. This may occur only for $1+z>55(\Omega_{\rm
IGM}h_{50}^2/0.05)^{-1/2}$. At $z\sim 6$, collisions will mix the
hyperfine levels if $\delta\rho/\rho\gsim 100$, i.e., in the case of nearly
virialized halos. 

\subsection{Radio Loud Quasars}

So far, we have limited our attention to radio quiet sources of radiation.
However, 21-cm radiation emitted by quasi-stellar radio sources could raise
the spin temperature of neighboring intergalactic \HI above the CBR temperature
(Bahcall \& Ekers 1969; Urbaniak \& Wolfe  1981). At a distance $r$ from the
QSO, the brightness temperature at 21-cm is
\begin{equation}
T_{R}={c^2\over 8\pi k\nu_{10}^2}{L_\nu\over 4\pi r^2},
\end{equation}
where $L_\nu$ is the absolute luminosity of the radio-emitting QSO at 
1.4$\,$GHz. The typical radio quasar power is a small fraction of the B-band 
luminosity,
\begin{equation}
(\nu L_\nu)_{5\,{\rm GHz}}\approx 4\times 10^{-4} (\nu L_\nu)_{B}
\end{equation}
(e.g., Phinney 1985). So, assuming $\nu L_\nu$ is roughly constant in the 
radio, we derive
\begin{equation}
T_{R}\approx (30\,{\rm K})\left[{(\nu L_\nu)_B\over 10^{47}\,{\rm ergs~
s^{-1}}} \right]r_{\rm Mpc}^{-2}.
\end{equation}
This temperature must be added to $T_\CBR$ in equation~(\ref{eq:tspin}) for
determining the level populations. Comparison with
equation~(\ref{eq:plpalpha}),
however, shows that in the presence of UV photons this term is very small
compared to the \Lya pumping term and so can safely be neglected.

\section{Preheating of the IGM}

While a cold IGM may be detected in absorption against the CBR,
when the spin and kinetic temperatures
are coupled through the Wouthuysen-Field mechanism, the period for which
absorption is possible is short-lived. The reason is that the \Lya photons
that couple the spin and kinetic temperatures will quickly heat the IGM through
atomic recoil to a temperature above that of the CBR. In this section, we
return to the radiation sources introduced in \S~2 and discuss their roles as
heating sources by \Lya scattering and by photoelectric heating by soft X-rays.

\subsection{\Lya Heating}

The effect of recoil due to the finite mass of the atom was first
discussed in resonance-line scattering by Field (1959b) (see also Adams 1971;
Basko 1981). Contrary to the case of radiation scattered by free electrons
(where a systematic shift in frequency due to recoil can produce large
cumulative effects), in problems of resonance-line transfer a photon on the red
side of the line has a heavy bias toward scattering back to the blue. This bias
suppresses the additive effect of recoil: because of Doppler redistribution,
a photon largely loses memory of its injection energy, and the background 
intensity develops only a slight asymmetry in frequency at line
center (matching the slope of the Planck function in the Wien limit, see Field
1959b) even at the large \Lya optical depths normally encountered in
astrophysical situations. However, it is through the recoil of the atom that
the radiation field exchanges energy with the gas in scattering problems. In
this section we shall discuss in some detail the effect of
repeated \Lya resonant scattering on the thermal state of a cold, neutral IGM.
A more formal derivation of the rate of transfer of energy from the radiation
field to the atoms is provided in Appendix B. 

The average relative change in a \Lya photon's energy $E$ after having been
scattered by a hydrogen atom at rest is
\begin{equation}
\langle{\Delta E\over E}\rangle=-{h\nu_{\alpha}\over m_\nH c^2}\approx 
-10^{-8}, \label{eq:de_e}
\end{equation}
where $m_\nH$ is the mass of the hydrogen atom. It should be noted that this is
an approximation valid only for $h\nu_\alpha\gg kT_K$. In the opposite limit,
energy will flow from the atoms to the photons. Through recoil, energy is
transferred from photons to atoms at a rate 
\begin{equation}
\dot E_\alpha=-\langle{\Delta E\over E}\rangle h\nu_{\alpha}
P_\alpha. \label{eq:edot}
\end{equation}
where $P_\alpha$ is the \Lya scattering rate per H atom. In the case of
excitation at the thermalization rate $P_{\rm th}$, equation~(\ref{eq:edot})
becomes
\begin{equation}
\dot E_{\rm th}={{27}\over4}{{(h\nu_\alpha)^2}\over{m_\nH c^2}}
{{A_{10}T_{\rm CBR}}\over{T_*}} \approx
(220~{\rm K\,Gyr}^{-1})\left({{1+z}\over7}\right). \label{eq:Ethdot}
\end{equation}
{\it This rate is sufficient to drive the intergalactic gas kinetic temperature
above the temperature of the cosmic background radiation in a fraction $\approx
15\%$ of the Hubble time at $z=6$.} As a consequence, a low density IGM is
observable in absorption at 21-cm for only a brief interval of time once
coupling is established. In general, coupling through the Wouthuysen-Field
mechanism leads to emission. 

This may be illustrated for the case of an early population of stars. It
follows from equations~(\ref{eq:Jast}) and (\ref{eq:pj}) that
the equivalent of only $\sim 10\%$ of the global present-day star formation
rate is required for $P_\alpha$ to exceed $P_{\rm th}$, and so drive
$T_S\rightarrow T_K$. The IGM would initially be observed in absorption, but
after $\sim 10^8\,$yr, the IGM will be heated to a temperature above
$T_{\rm CBR}$ by resonant scattering of the \Lya photons.

It is of interest to compare \Lya heating with heating by X-ray
photons.  The amount of energy deposited via photoionization by soft
X-rays is, allowing for secondary ionizations, a few tens of eV per
photon. There are also a few free electrons for every photon, as
recombinations are negligible. In contrast, the total energy available
for \Lya heating (whether the \Lya photon is part of the continuum or
due to recombinations) is, neglecting the energy lost to cosmological
expansion, given by the photon energy multiplied by the fractional
line width.  This is typically 0.001 eV per photon. Therefore, at
least of order one \Lya photon per atom is needed to raise the kinetic
temperature of the IGM above $T_\CBR$.  If we denote by $L_X$ the soft
X-ray luminosity of a young galaxy (associated with, e.g., supernova
remnants, X-ray binaries, etc.), then photoelectric heating will
dominate over \Lya heating if $L_X/L_\alpha\gsim 10^{-4}$. As the
typical observed value for $L^\star$ galaxies is $L_X/L_\alpha\sim
10^{-2}$ (e.g., Fabbiano 1989; Bennet \etal 1994), \Lya photons from
galaxies are an inefficient heating source compared to X-rays.
\footnote {As shown by Rees (1985), the radiation pressure associated
with trapped \Lya background photons is also unlikely to affect the condensing
out of overdense regions of the IGM even in the linear regime.} 

\subsection{Soft X-ray Heating}

\subsubsection{Heating Rate by Quasars}

It is widely believed that the integrated ultraviolet flux arising from
QSOs and/or hot, massive stars in metal-producing young galaxies is responsible
for maintaining the intergalactic diffuse gas and the \Lya forest clouds in a
highly ionized state at $z\lsim 5$. In this and the next section we shall
focus on a reionization scenario dominated by hard-spectrum quasar sources. 
In this case, while the radiation just shortward of the hydrogen Lyman edge will
be absorbed at the ionization front generated by an individual QSO, very short
wavelength photons will be able to propagate to much greater distances,
$r_c>r_I$. The intensity of the QSO radiation flux at the light front will be
reduced by the opacity of the (mostly) neutral gas surrounding the expanding
\HII region,
\begin{equation}
\tau_\nu = n_\nHI(r_c-r_I)\left(\sigma_\nu^\nHI + \chi\sigma_\nu^{\rm
HeI}\right), \label{eq:taueq}
\end{equation}
where $\sigma_\nu^\nHI$ and $\sigma_\nu^\nHeI$ are the hydrogen
and helium photoionization cross sections. For $\Omega_{\rm IGM}h_{50}^2=0.05$,
$z>5$, and $r_c-r_I=10\,$Mpc, the photoelectric optical depth will drop below
unity for photon energies above 16 ryd. Thus, most of the photoelectric heating
of the IGM will be accomplished by soft X-rays. The heating by such high energy
photons, however, is not as efficient as for photons near the photoelectric
threshold because of losses to collisional excitation of \HI and \HeI and
to the collisional production of secondary electrons (Spitzer \& Scott 1969).
The fraction $f$ of photon energy converted into heat is then a
sensitive function of the \HII fraction. At the light front, this fraction is
the residual ionization left over from the recombination epoch. For $h_{50}=1$
and $\Omega_{\rm IGM}=0.05$, this is $\sim5\times10^{-4}$ (Peebles 1993), which
gives $f\approx0.17$ (Shull \& van Steenberg 1985). For a quasar with a
spectral index $\alpha_S$ shortward of 1$\,$ryd, the heating rate per hydrogen
atom at the light front is given by
\begin{equation}
\dot E_X(r_c)=\alpha_S {S\over 4\pi r_c^2}h\nu_L f
\int_1^\infty dx x^{-1-\alpha_S} [\sigma^{\rm HI}_\nu
(x-1) + \chi\sigma^\nHeI_\nu (x-1.8)]e^{-\tau_\nu},
\end{equation}
where $x\equiv\nu/\nu_L$, and $\nu_L$ is the frequency at the \HI Lyman
edge. Both the \HI and \HeI photoelectric cross sections scale approximately
as $\nu^{-3}$ far from threshold. In this approximation the integral becomes
exact, and may be expressed as
\begin{eqnarray}
\dot E_X(r_c) \approx~ & \left[(1400 ~{\rm K\, Gyr}^{-1})\alpha_S
\Gamma\left({{2+\alpha_S}\over3}\right)
\tau_*^{(1-\alpha_S)/3}\right]\left({f\over{0.17}}\right) \left({S\over
10^{57}\,{\rm s}^{-1}}\right)\left( {\Omega_{\rm IGM}h_{50}^2\over
0.05}\right)^{-(2+\alpha_S)/3} \nonumber \\ & \times \left({1+z\over
7}\right)^{-2-\alpha_S}\left(1-{r_I\over r_c}\right)^{-(2+ \alpha_S)/3}
\left({{r_c}\over{25\mpc}}\right)^{-(8+\alpha_S)/3},
\end{eqnarray}
where $\Gamma(x)$ is the usual gamma function. Here, $\tau_*$ normalizes
the opacity of the IGM for soft X-ray frequencies according to $\tau_\nu=
\tau_* (\Omega_{\rm IGM}h_{50}^2/0.05)[(1+z)/7]^3(r_c/25\mpc)(1 - r_I/ r_c)
(\nu/ \nu_L)^{-3}$. Numerically, $\tau_*\approx 7.8\times10^4$. For
$\alpha_S=1.5$, the coefficient in square brackets is $\sim 300 ~{\rm K\,
Gyr}^{-1}$. The heating rate is dominated by helium absorption, which exceeds
the hydrogen contribution by $\sim 2$:1. The intergalactic gas at the light
front will be heated to a temperature above that of the cosmic background
radiation in a fraction of the Hubble time, 
\begin{eqnarray}
{\Delta t_{\rm heat}\over t_H}\approx~ & \left[0.02 {{\tau_*^
{(\alpha_S-1)/3}} \over{\alpha_S \Gamma\left\{(2+\alpha_S)/3\right\}}}\right]
h_{50}\left({{f}\over{0.17}}\right)^{-1}
\left({S\over10^{57}\, {\rm s}^{-1}
} \right)^{-1}\left({\Omega_{\rm IGM}h_{50}^2\over0.05}\right)^ {(2+\alpha_S)/3
} \nonumber \\ & \times\left({1+z\over 7}\right)^{9/2+\alpha_S} \left(1-{r_I\over r_c}
\right)^{(2+\alpha_S)/3} \left({{r_c}\over{25\mpc}}\right)^{(8+\alpha_s)/3}.
\end{eqnarray}
For $\alpha_S=1.5$, the coefficient in square brackets is $\sim0.1$. The \HII
region produced by a QSO will therefore be preceded by a warming front. Heating
by soft X-rays is so efficient that the size of the warm zone will extend
nearly to the light front. Note that, as the X-ray-heated bubbles around QSOs
will survive as fossils even after the quasar has died, several generations
$N_g$ of quasars may actually be responsible for preheating the entire IGM. For
a typical QSO age of $t_Q\approx3\times10^8 N_g^{-1/3}$ yr, the required QSO
comoving space density to heat the entire IGM to a temperature above that of
the CBR by $z\approx 6$ is $\sim10^{-10}\mpc^{-3} N_g^{-1}$. By comparison, the
comoving space density of bright QSOs at $z=4$ is $\sim 100N_g$ times larger
(Warren, Hewett, \& Osmer 1994). Note that, if all bright galaxies undergo a
quasar phase, QSOs  must have a very short lifetime, and $N_g\sim 100$. Soft
X-rays from a few bright QSO sources could then prevent collapsing structures,
such as protoclusters while still in the linear regime, from being detected in
21-cm absorption against the CBR (Scott \& Rees 1990). 

\subsubsection{Numerical Solution for Expanding Warm \HI Regions
Around Quasars}

We illustrate the thermal effect of an expanding \HII region on the
surrounding medium by numerically solving the time-dependent equations
for photoionization and heating around a point source which turns on
suddenly in a homogeneous, expanding universe. We have integrated as a
function of time the ionization equations for hydrogen and helium,
including photoionization, collisional ionization, and radiative
recombination, on a radial grid with a local ionization flux equal to
$L_\nu\exp(-\tau_\nu)/4\pi r^2$, where $\tau_\nu$ is given by
equation~(\ref{eq:taueq}). The flux is set to zero beyond the light
front to prevent unphysical excess heating at large distances. We use
the time-dependent ionization algorithm of Meiksin (1994), and include
here the generation of secondary electrons by soft X-rays. We adopt
the fitting formulas of Shull \& van Steenberg (1985) to include the
additional ionization and energy losses associated with the production
of these electrons. We solve the time-dependent energy equation along
with the ionization equations for a gas of primordial hydrogen and
helium composition, including photoelectric heating and radiative
losses due to collisional excitation and ionization, radiative
recombination, Compton cooling off the CBR, and cooling by expansion.
An accurate numerical solution for the growth of the \HII region
requires grid zones that are optically thin to the ionizing radiation
when the hydrogen is still fully neutral. Thicker zones are ionized
too rapidly, resulting in a too rapidly growing I-front. Because we
wish to follow the heating out to the light radius, however, this
restriction imposes a prohibitive demand on our computing resources.
The demand imposed by helium ionization is less restrictive because of
its lower density, and so may be met. We therefore adopt a compromise.
A grid sufficiently fine to resolve the helium ionization is used, but
it is somewhat too coarse to follow the evolution of the \HII I-front
accurately. To prevent too rapid a growth of the \HII front, we set
the flux to zero for energies less than 1.8 ryd, the \HeI ionization
potential. We find this results in accurate placement of the \HII
front, though the details of its structure may not be quite correct.
Because helium ionization and the soft X-ray heating are governed by
the opacity outside the \HII front, this approach yields an accurate
solution for the \HeII and \HeIII fronts and the external heating rate
in the surrounding neutral IGM.

In Figure 1, we show the results of an integration for a QSO turning on at
$z=6$ with a gradual rise over a time of $10^7\,$yr to a peak photoionizing
luminosity of $S=10^{57}\,$s$^{-1}$, and with $\alpha_S=1.5$, in a medium with
$\Omega_{\rm IGM}h_{50}^2=0.05$. By $z=5.5$,
the QSO has heated the surrounding gas to a temperature exceeding that of
the CBR to a (proper) distance of 23 Mpc, compared to the light front radius of
25 Mpc. The temperature climbs inward, reaching $1000\kel$ at a distance of 10
Mpc from the QSO. The ionization fronts are at 7.5 Mpc at this time. Neither
the \HeII nor the \HeIII fronts extend much beyond that of hydrogen, but for
different reasons. Because of the much higher photon energy required to 
photoionize \HeII compared to \HI, the radius of the \HeIII front can exceed
that of \HII by at most a factor of $(4^{-\alpha_S}/\chi)^{1/3}\simeq 1.1$, 
despite the low helium abundance (cf. Madau \& Meiksin 1994). While the energy
needed to photoionize \HeI is smaller, the IGM is more optically thick at the
lower energy. At $z=5.5$, the opacity of \HI at the \HeI edge reaches unity a
distance 10 kpc beyond the \HII ionization front. Outside the \HII region, 
\HI efficiently shields the helium from the ionizing radiation
above 1.8 ryd, ensuring that it remains mostly neutral. The \HeI is then able
to warm through the absorption of the incident soft X-rays emitted by the QSO. 

\subsubsection{Heating Rate by Virialized Halos}

Galactic mass halos provide another source of heating in CDM-dominated
cosmologies. While before stars form the numbers of these halos are small,
too small to account for the ionization of the IGM by UV photons, we
demonstrate here that a sufficient number of halos will collapse to provide
the soft X-rays required to heat the IGM to a temperature above that of
the cosmic background radiation. 

We start with equation~(\ref{eq:ehalo}), which gives the background emissivity
above a minimum photon energy. To specify its value, we observe that,
at sufficiently low energies,
the distance a photon travels before being absorbed in the IGM will be shorter
than the mean separation of the sources, so that only part of the IGM will be
penetrated by such photons. We set the minimum energy according to
the criterion that the filling factor of photons with energy above
$h\nu_{\rm min}$ be greater than unity.  For $5<z<10$, this yields
\begin{equation}
h\nu_{\rm min}\approx 15.2\, {\rm ryd}\left({{1+z}\over7}\right)^{3/2},
\label{eq:numin}
\end{equation}
corresponding to a halo of comoving size $\approx0.9\mpc$, 
virial temperature $\approx1.5\times10^6\kel$, and total mass $\approx 
2\times10^{11}\msun$.
Approximated in this manner, the emissivity is only half of the value given
in equation~(\ref{eq:ehalo}), since halos with $r_0<\hat r_0$ are excluded by
the
filling factor criterion. Note that the collisional ionization level of helium
in such hot halos is so large that absorption of soft X-rays within the halos
themselves is completely negligible. Substituting equation~(\ref{eq:numin})
into equation~(\ref{eq:ehalo}), and taking into account the factor of 2, we
find for the heating rate of the halos per hydrogen atom
\begin{equation}
\dot E_X^{\rm halo}\approx~ (2.6\times10^5\, {\rm K\, Gyr^{-1}})
\left({f\over{0.2}}\right)
\left({{1+z}\over7}\right)^3 \exp\left[-6.6\left({{1+z}\over7}\right)^{1.6}
\right], \label{eq:ehhalo}
\end{equation}
where $f\approx0.2$ accounts for the energy loss to collisional excitations and
ionizations per incident X-ray photon (Shull \& van Steenberg 1985), for an
assumed \HII fraction of $10^{-3}$. We show the heating rate in Figure 2. The
IGM will be heated to a temperature above that of the CBR in a fraction of the
Hubble time,
\begin{equation}
{{\Delta t_{\rm heat}^{\rm halo}}\over t_H}\approx~ (7.5\times10^{-5})
\left({f\over{0.2}}\right)^{-1}
\left({{1+z}\over7}\right)^{-0.5}\exp\left[6.6\left({{1+z}\over7}\right)^{1.6}
\right]. \label{eq:thheat}
\end{equation}
At $z=6$, this fraction is 0.05. Because the merging of small halos into
larger ones is an ongoing process, the full Hubble time will not be available 
to the halos at any given stage in the hierarchy for heating the IGM. Moreover,
the halos will cool by radiative processes on a timescale of $t_{\rm cool}/
t_H\simeq0.7[(1+z)/7]^{-3/2}T_6^{1.8}$ (Gould \& Thakur 1970),
and even more quickly at high redshifts by Compton cooling off the CBR. The
cooling, however, may not be significant for $z<10$ before the gas is reheated
by mergers in the next stage of the hierarchy. Allowing the halos to provide
X-ray photons for 50\% of the Hubble time, they will heat the IGM to a
temperature of $\sim100\kel$ by $z=6$. At higher redshifts, halo heating may
still be important in overdense environments like protoclusters where the
number density of halos may be substantially boosted by statistical bias
(Bardeen et al. 1986). Of course once stars start to form, they or, more 
likely, their remnants may dominate the soft X-ray emission of an individual
halo. In this sense, the halo emissivity calculations outlined above present
a lower limit to the heating rate of the IGM at early epochs.

It is worth noting that while the X-rays are an efficient heating mechanism of
the gas, they are an inefficient source of photoionization. This is because the
time required to photoionize the IGM by X-ray photons is larger by a factor of
$\sim (15\, {\rm ryd}/ kT_\CBR)\approx10^5$, since now it is the number of
ionizing photons rather than their energy that counts. The ionization time thus
exceeds the Hubble time by a few orders of magnitude, and the IGM will remain
mostly neutral.

\subsubsection{Thermal Coupling Between Neutral Atoms and Ions}

In the discussion above, we have assumed that the energy deposited by the
photons is strongly coupled to the hydrogen. The thermal energy, however,
is initially contained in the photoejected electrons and ions only.
In this section we shall verify that the equilibration mechanisms between
the various ions and neutrals in the low density IGM are sufficiently rapid
to ensure that all species are thermally coupled.

We first consider Coulomb coupling between the ions. The time for two different
species $i$ and $j$ to reach thermal equilibrium through Coulomb scattering is
\begin{equation}
t_{\rm eq}(i,j)\approx (5.87\,{\rm s}){{A_iA_j}\over{(n_i+n_j)Z_i^2Z_j^2
\log\Lambda}} \left({T_i\over A_i} + {T_j\over A_j}\right)^{3/2},
\label{eq:teqeq}
\end{equation}
(Spitzer 1962). Each species has a temperature $T$, density $n$, charge
$Z$, and molecular weight $A$ ($A=1/1836$ for electrons).
(This expression applies strictly only when the total energy between the
two species, $n_iT_i+n_jT_j$, is conserved during the equilibration process.
For a single species, $j=i$ may be set in equation~(\ref{eq:teqeq}) if the
coefficient 5.87 is replaced by 8.06.) In the physical regime of interest, 
$20<\log\Lambda<50$.
Once the electrons and protons each reach equilibrium separately (the
shortest two timescales), the bottleneck for all the ions to relax to a single
temperature is proton-electron scattering, with a coupling time
\begin{equation}
t_{\rm eq}(e,p)\approx (3.3~{\rm Myr}) \left({{x_e}\over{10^{-3}}}\right)^{-1}
\left({{1+z}\over7}\right)^{-3}
\left({{\Omega_{\rm IGM}h_{50}^2}\over{0.05}}\right)^{-1}T_4^{3/2},
\end{equation}
where $x_e$ is the number of electrons per hydrogen nucleus.  After these 
two species equilibrate, singly and doubly ionized helium follow shortly 
through proton scatters. 
Note that, as the post-photoionization (by 16 ryd soft X-rays) temperature of 
electrons is $\sim 10^6\,$K, the
time for protons to thermalize with such high energy electrons by Coulomb
collisions alone actually exceeds the Hubble time. At these energies, however, 
the primary photoelectrons will quickly cool via the production of 
secondary electrons
by the collisional ionization and excitation of neutral hydrogen
(Spitzer \& Scott 1969). The cooling rates of an electron of energy $E$ ryd by
these two processes are comparable, and are given by
\begin{equation}
t_{\rm cool}\approx (0.03~{\rm Myr}) \left({{1+z}\over7}\right)^{-3}
\left({{\Omega_{\rm IGM}h_{50}^2}\over{0.05}}\right)^{-1}
{{E^{3/2}}\over{\log E}},
\end{equation}
(Shull \& van Steenberg 1985). For $E=16$, this gives $t_{\rm cool}\approx0.6$
Myr. The result is an energy distribution of primary and
secondary electrons which declines sharply above the \Lya excitation energy of
10.2 eV (Bergeron \& Souffrin 1971).

The coupling rate is still small for electrons of these energies,
but they further lose energy through elastic scattering  off neutral
hydrogen atoms. The electron-hydrogen cross section for a 10 eV electron is
about $7\pi a_0^2$, where $a_0$ is a Bohr radius (Moiseiwitsch 1962). The
corresponding relaxation timescale is
\begin{equation}
t_{\rm eq}(e,{\rm H})\approx (25~{\rm Myr}) \left({{1+z}\over7}\right)^{-3}
\left({{\Omega_{\rm IGM}h_{50}^2}\over{0.05}}\right)^{-1}. \label{eq:teHeq}
\end{equation}
The electrons can finally cool to temperatures below $\sim10^4\,$K.
Once a thermal distribution of electrons has been established, new energetic
electrons generated by X-ray photoionization will reach equipartition with
the cooler electrons on a much shorter timescale by Coulomb losses
(Shull \& van Steenberg 1985).

The distance from the light-front for which the temperature of the neutral
hydrogen atoms lags behind the electron temperature corresponding to
equation~(\ref{eq:teHeq}) is $\sim 8\mpc$. This, however, assumes complete
thermalization.
The less stringent requirement that the neutral hydrogen atoms be heated to a
temperature large compared to the cosmic background radiation temperature will
be satisfied on a somewhat shorter timescale. The number of QSOs required to
heat the IGM above the cosmic background radiation temperature may then 
exceed the estimate given in \S~4.2.1 by a factor of a few only.

The neutrals will reach a single common temperature relatively quickly. The
interaction between neutral hydrogen atoms is dominated by elastic scattering.
\footnote{Provided that statistical equilibrium has been established between
the singlet and triplet ground state levels. This is achieved on a timescale 
$\sim 5\times 10^4\,$yr by collisions with CBR photons.}~ 
The equilibration time is
$t_{\rm eq}=2/n_\nHI\langle u\sigma\rangle$, where $\langle u\sigma\rangle$
is the slowing down coefficient for hydrogen atom collisions (Spitzer 1978).
Using the elastic collision cross section from Allison \& Smith (1971),
this gives
\begin{equation}
t_{\rm eq}({\rm H,H})\approx (0.4~{\rm Myr}) \left({{1+z}\over7}\right)^{-3}
\left({{\Omega_{\rm IGM}h_{50}^2}\over{0.05}}\right)^{-1}T_4^{-1/4}.
\end{equation}
Thus all the species will reach a common temperature within a timescale
much shorter than the Hubble time, and it is safe to assume that the hydrogen
atoms may be described by a single Maxwellian energy distribution.

\section{Observations of Redshifted 21-cm Line Radiation}

In this section we discuss how radio measurements may reveal 21-cm line
radiation from a neutral IGM at high redshift. The {\it Giant Metrewave Radio
Telescope} (GMRT), scheduled for completion by 1997, has two frequency bands
which are best suitable for detecting 21-cm radiation at $z>5$, viz.
$150\pm 20$ and $235\pm 8\,$MHz (Swarup 1996). The central frequencies of
these bands correspond to $z=8.5$ and 5.0, respectively. Because of its high
sensitivity for both compact and extended sources, the GMRT will be a valuable
instrument for cosmological studies in the redshift ranges $4.85<z<5.25$ and
$7.35<z<9.9$. We focus the discussion on measurements in the 150 MHz band,
corresponding to $z\approx8$. 

We have assumed that the IGM at early epochs is uniform on large scales. One
general issue is the effect of small-scale clumping. In a CDM-dominated
cosmology,
inhomogeneity on subgalactic scales develops before the first quasars form
(if the latter require galactic-mass collapsed structures). This may delay the
eventual percolation of \HII regions, and increases the value of $J$ needed to
balance recombinations and maintain full ionization. It is unlikely that
clumping would change the present discussion by very much. It leaves unchanged
the amount of heating and ionization due to each UV photon, as well as the rate
at which the heating and ionization fronts advance around a quasar. Even in an
overdense region (unless the overdensity is extreme), recombinations are
unimportant over the lifetime of a single quasar (or during the time it takes
for the heating front to advance). It is also worth emphasizing that there will
unavoidably be nonuniformities in the 21-cm intensity due to density
inhomogeneities, but that these are restricted to smaller scales than those due
to the patchy heat input (at least in the QSO case). While we confine our
discussion to larger scales, it would nonetheless be worthwhile to consider
means for their detection.

To illustrate the basic principle of the observations we propose, consider a
region of neutral material with spin temperature $T_S\neq T_\CBR$, having
angular size on the sky which is large compared to a beamwidth, and radial
velocity extent due to the Hubble expansion which is larger than the bandwidth.
Its intergalactic optical depth at $21(1+z)\,$cm along the line of sight, 
\begin{equation}
\tau(z)={3c^3 h^3 n_\nHI(0)A_{10}\over 32\pi H_0 k^3 T_*^2 T_S}(1+z)^{1.5}
\approx 10^{-2.9}h_{50}^{-1}\left({T_\CBR\over T_S}\right)
\left({\Omega_{\rm IGM}h_{50}^2 \over 0.05}\right) (1+z)^{1/2},
\end{equation}
will typically be much less than unity. The experiment envisaged consists of
two measurements, separated in either angle or frequency, such that one
measurement, the fiducial, detects no line feature, either because there is
no \HI or because $T_S\approx T_\CBR$, and the second at $T_S\neq T_\CBR$. 
As the brightness
temperature through the IGM is $T_b=T_\CBR e^{-\tau}+T_S(1-e^{-\tau})$, the
differential antenna temperature (which is the relevant quantity since we are
considering a detector operating as a comparison system), observed at Earth
between this region and the CBR  will be
\begin{equation}
\delta T=(1+z)^{-1} (T_S-T_\CBR) (1-e^{-\tau})\approx (0.011~{\rm K})
h_{50}^{-1} \left({\Omega_{\rm IGM}h_{50}^2\over 0.05}\right) \left({{1+z}\over
9}\right)^{1/2} \eta, \label{eq:dT}
\end{equation}
where we have defined a ``21-cm radiation efficiency'' as 
\begin{equation}
\eta\equiv x_\nHI\left({T_S-T_\CBR\over T_S}\right). \label{eq:eff}
\end{equation}
Here $x_\nHI$ refers to the neutral fraction of the hydrogen in the region
for which $T_S\neq T_\CBR$. As long as $T_S$ is much larger
than $T_\CBR$ (hence if there has been significant preheating of the
intergalactic gas), $\eta\rightarrow x_\nHI$, and the IGM can be observed in
emission at a level which is independent of the exact value of $T_S$. By
contrast, when $T_\CBR\gg T_S$ (negligible preheating), the differential
antenna temperature appears, in absorption, a factor $\sim T_\CBR/T_S$ larger
than in emission, and it becomes relatively easier to detect intergalactic
neutral hydrogen (Scott \& Rees 1990). Note that in a universe with 
$q_0\ll 1/2$, $\delta T$ increases more nearly linearly with $(1+z)$, and the 
numerical coefficient in equation~(\ref{eq:dT}) may be larger by up to a factor
$3[(1+z)/9]^{1/2}$. Depending on the cosmology, $\delta T$ will increase toward
low frequencies at a rate between $\nu^{-1/2}$ and $\nu^{-1}$. 

The discrete nature of the sources of preheating and 21-cm excitation will
cause $\delta T$ to have a distinctive structure both in frequency and in
angle across the sky, as will be shown below.
It is important to point out that the radio signal from
the IGM will be swamped by the much stronger non-thermal background that
dominates the radio sky at meter wavelengths. This synchrotron-type component 
has brightness temperature
\begin{equation}
T_b(z=0)\approx 6\kel\left({\lambda\over{1\,{\rm m}}}\right)^{2.8},
\end{equation}
which is consistent with the integrated contribution from discrete radio
sources (see Longair 1995 for a recent review). The anisotropic galactic 
contribution is about three times larger.  Such  emission will cause 
intensity fluctuations from beam to beam, either from discrete radio sources
or from variable galactic brightness. Hence the difficulty is to be able to 
discriminate between the IGM around a QSO with a brightness temperature of 
$0.01\kel$ against a small fluctuation in the galactic and extragalactic 
background. Clearly, to remove this foreground contamination,
some differential mapping technique must be used. The line radiation from 
high-$z$ gas might, e.g., be distinguished against continuum emitting sources
owing to its sharp and distinctive spectral feature, i.e., by making 
measurements at two frequencies sufficiently close that the continuum
emission from discrete sources does not vary significantly (Bebbington 1986).

In the following sections we will illustrate by a few examples the sensitivity
of the detailed pattern of sky brightness at redshifted 21-cm wavelengths to
the unknown thermal and ionization history of the universe. 

\subsection{Quasars in a Background of UV Radiation at ${\bf z\approx 8}$}

In the theories for the origin of cosmic structure that are most popular today,
massive objects grow by hierarchical clustering -- low mass perturbations
collapse early, then merge into progressively larger systems. In such a
scenario, one naturally expects the high redshift universe to contain small
scale substructures, systems with virial temperature well below $10^4\,$K. The
resultant virialized systems will remain as neutral gas clouds unless they can
cool due to molecular hydrogen, in which case they may form Pop III stars.
These stars will then ionize the gas in their vicinity. It is less clear,
however, whether or not a sufficient number of ionizing photons will be able to
penetrate all of the surrounding neutral gas in the cloud to reach the external
intergalactic gas and ionize it. Nonetheless, the stars will still produce a
background of UV continuum photons near the \Lya frequency which can escape.

It is therefore of interest to consider a case in which QSOs turn on in a
universe where the \Lya scattering rate is $P_\alpha\approx P_{\rm th}$
everywhere, i.e., where
\begin{equation}
J_{\alpha}=J_{\rm th}\equiv {9\over 2}{A_{10}\nu_\alpha^3\over
A_\alpha\nu_{10}} {kT_\CBR\over c^2}\approx (10^{-21.1}\uvunits)
\left({1+z\over 9}\right), \label{eq:Ja}
\end{equation}
(cf. eqs. [\ref{eq:ptherm}] and [\ref{eq:pj}]). The \Lya photons will
propagate into
uncollapsed regions of the IGM where the kinetic temperature in the absence of
heating will be
\begin{equation}
T_K(z)\approx 0.026\kel (1+z)^2 \label{eq:Tc}
\end{equation}
(Couchman 1985), well below $T_\CBR$ because of adiabatic cooling during cosmic
expansion. The \Lya flux will couple the spin temperature to
the kinetic temperature, and $T_S$ will be pulled below $T_\CBR$. (Note that,
in the absence of a radiative coupling agent like \Lya, this could only happen
in very high-baryon-density models, where collisions would remain important
even at modest redshifts, Scott \& Rees 1990.)

As shown in \S~4.1, however, the same \Lya photons which mix the hyperfine 
levels will also heat the gas through recoil. Integrating the energy
equation, assuming the \Lya field is turned on, or more precisely reaches
the level of equation~(\ref{eq:Ja}), at redshift $z_\alpha$, and
using equations~(\ref{eq:Ethdot}) and (\ref{eq:Tc}), we find
\begin{equation}
T_K(z)\approx (0.026 \kel)(1+z)^2+ (160\kel)(1+z)^2[(1+z)^{-5/2}-(1+z_\alpha)
^{-5/2}]. \label{eq:TM}
\end{equation}
The characteristic timescale for heating the medium above the CBR temperature
via \Lya resonant scattering at the thermalization rate is
\begin{equation}
\Delta t_{\rm heat}={2\over 9} {m_\nH c^2\nu_{10}\over h\nu_\alpha^2}
A_{10}^{-1}\approx 10^8\,{\rm yr},
\end{equation}
about 20\% of the Hubble time at $z\approx8$. The heating rate corresponding to
equation~(\ref{eq:Ja}) is shown in Figure 2. The result is a finite interval
of time during which \Lya photons couple the spin temperature to the kinetic
temperature of the IGM before heating the IGM above the CBR temperature.
If \Lya sources turned on at redshifts $z_\alpha\lsim 10$, this interval would
present a window of opportunity in redshift space near $z\approx8$ that would
enable a large fraction of intergalactic gas to be observable at
$\sim 160\,$MHz in {\it absorption} against the CBR.

In such a scenario, 21-cm emission on tens of megaparsecs scale would still be
produced by the relatively rare, X-ray heated regions associated with QSO
sources. If we denote by $T_{S,w}$ the spin temperature of the ``warm''
($T_{S,w}\gg T_\CBR$) gas, and by $T_{S,c}$ the spin temperature of the
surrounding, more pervasive ``cold'' ($T_{S,c}<T_\CBR$) medium, as given by
equation~(\ref{eq:TM}), the differential antenna temperature observed between
these two phases of the IGM will be proportional to the effective efficiency
\begin{equation}
\Delta\eta=T_\CBR\left({x_{\nHI,c}\over T_{S,c}}-
{x_{\nHI,w}\over T_{S,w}}\right). \label{eq:deta2}
\end{equation}
Here, we distinguish between the neutral fractions in the cold and warm
regions, to allow the warm zone to refer either to the inside or the outside
of the ionized bubble. Even though the \HI in the cold region is absorbing
the CBR, by using the flux through the cold gas as a fiducial, the signature of
the cold gas will appear as {\it emission} from the warm regions surrounding
the quasars. The effective efficiency will remain much larger than unity,
$\Delta\eta\approx T_\CBR/T_{S,c}\gsim 4$, only for $\lsim 10^7\,$yr
(cf. equation~[\ref{eq:TM}]), to drop to $\Delta\eta\approx 2$ after
$\approx 5\times 10^7\,$ yr. Note that $\Delta\eta\approx 4$ corresponds to
temperature fluctuations of about $0.04\kel$. This flux is well below the level
of the Bebbington's (1986) radio survey, but would be reachable with an
instrument like the GMRT in several hours of integration.

For illustrative purposes, consider now the sphere of influence associated with
a quasar ($L_{\alpha,47}=1$, $\alpha_S=1.5$) which turned on at
$z=8.7$, nearly synchronized with a \Lya background of sources turned on at
$z=9$, and radiating isotropically.
At $z=8$, e.g., after $5\times 10^7\,$yr the QSO will have ionized 
a bubble of size $r_I\approx 7.5\mpc$, and heated the surrounding \HI within 
$r_c\approx 15\mpc$. In Figure 3a we show $\Delta\eta$ as a function of radius
at various epochs, including the effect of \Lya heating on the cold phase. The
peak value at the I-front initially exceeds unity, but declines as
$T_{S,c}$ rises. The quasar has created a shell of 21-cm
emission surrounding the \HII region with a thickness of $\Delta r
\approx7.5\mpc$.  The hydrogen column density through the shell's width is
$\approx 2\times10^{21}\cmm$, and its total \HI mass is
$\approx 2\times 10^{16}\msun$. The angular size on the sky corresponding to
the diameter of the emitting region is 
\begin{equation}
2\theta={H_0 (1+z)^2r_c\over c(1+z-\sqrt{1+z})}\approx 2\,{\rm
degrees}.
\end{equation}
The shell will have a characteristic radial double peaked profile, as the 
gas in its interior will be completely ionized, as shown in Figure 3b.
The velocity difference across
each peak, due to the Hubble flow, is $\Delta v=H_0(1+z)^{3/2}\Delta r\approx
10,000\kms$. Between the inner edges of the two peaks, $\Delta v$ is
twice as large, since $\Delta r\approx r_I$ at this time. At $150\,$MHz, we
would then observe two signals (the front and back of the shell) with flux 
density
\begin{equation}
\delta S\approx (6~{\rm mJy\,}) h_{50}^{-1} \left({\Omega_{\rm
IGM}h_{50}^2\over 0.05}\right) \left({{1+z}\over 9}\right)^{-3/2}
\left({{\Delta\eta}\over 2.2}\right) \label{eq:dS}
\end{equation}
(cf. eqs. [\ref{eq:dT}] and [\ref{eq:deta2}]), extending over and separated by
about $5\,$MHz, 
nearly the entire bandwidth of the receiver.  For a fixed velocity width, 
$\delta S$ increases linearly with the mass of the shell.

Let us analyze in some detail the detectability of a typical 21-cm
emitting shell with the GMRT. The 30 antennas of the GMRT will be
configured as a Y of 16 dishes with a central compact core of 14
dishes. Because the GMRT functions as an interferometer, it is
insensitive to angular scales exceeding $\sim\lambda/d$, where $d$ is
the separation of the dishes. If we wish to measure $\sim 2$ degree-scale 
structures at $150\,$MHz ($\lambda\sim 2\,$m), the projected separation 
of antennas is required to be about $d\lsim50\,$m. Since the GMRT antennas have
a diameter of about 45$\,$m, the GMRT interferometric system will not be
sensitive to 2 degree-scale structures as the antennas will start shadowing
each other for projected spacings less than 45$\,$m.

An alternative strategy is to measure the total power by incoherent addition 
of the outputs of the 30 GMRT antennas, for which the effective collecting 
area will be $30\times 0.5$
that of one antenna. For observations at 150$\,$MHz the theoretical rms thermal
noise of the GMRT output in incoherent mode is $\approx 1\,$mJy for the 
following parameters: $T({\rm sys})\sim 600\,$K, $A({\rm eff})\sim 
1000\,$m$^2$ for 1 45-m dish, a bandwidth of $10,000\kms$ (5$\,$MHz), and 
5-hr integration for each of the 2 frequency bands separated by about 
5$\,$MHz (Swarup \& Malik 1996).  A comparison with equation~(\ref{eq:dS}) 
shows that $\sim 100\,$ hours of integration will be required to detect an
individual shell at 5$\sigma_{\rm rms}$. This time is comparable to
that proposed for searches for emission from collapsing \HI clouds at
high redshift (Swarup \etal 1991; Subramanian \& Padmanabhan 1993;
Swarup 1996).  However, it is not clear whether the above sensitivity can be
realized in practice because of the man-made noise and variations of the
instrumental gain and the galactic background temperature over 2 bands 
separated by 5$\,$MHz.
Note that, although the differential antenna temperature will be
larger at earlier epochs because of the larger brightness constrast
(i.e., less time to raise the temperature of the cold phase), the
smaller radius of the QSO sphere of influence more than cancels the
increase in emission efficiency. The signal-to-noise ratio improves
slightly with increasing bubble size. 

To calculate the expected number of emitting shells in a radio survey, we may
estimate their effective covering factor on the sky as 
\begin{equation}
C={c\over H}n_{\rm QSO}d_A^2(\theta+\Theta)^2{\Delta v\over c},
\end{equation}
where $n_{\rm QSO}$ is the (proper) number density of quasars, $d_A$ is
is the angular-diameter distance, $\Theta$ is the field of view of the
radio antenna (about 3.1 degrees at 150 MHz), and $\Delta v$ measures the
radial dimension covered by the survey. If the latter matches the shell
thickness $\Delta r$, e.g., for a bandwidth of $5\,$MHz, the covering factor
may be rewritten as
\begin{equation}
C=q_{21}\left(1+{\Theta\over\theta}\right)^2{\Delta r\over r_c}, \label{eq:cov}
\end{equation}
where $q_{21}$ is the volume filling factor of the 21-cm emitting bubbles. As
$\Theta^2\Delta r/\theta^2r_c\approx 5$, it follows from
equation~(\ref{eq:cov})
that a shell will be detected after scanning about 20 fields with the GMRT
if the filling factor of warm bubbles is $\approx 1$\%. 

\subsection{Isolated Quasars}

We shall now analyze the model problem of an individual, isolated QSO turning
on at epoch $z>5$ in an otherwise unperturbed IGM. In this case the only 
radiative coupling agent is the continuum UV radiation of the quasar itself,
redshifted to \Lya. From
equations (\ref{eq:plpalpha}) and (\ref{eq:ptherm}), it is useful to define a
``thermalization distance'' from a QSO, denoted by $r_{\rm th}$, at which
$P_\alpha=P_{\rm th}$. Numerically, this yields 
\begin{equation}
r_{\rm th}\approx (6\mpc) \left({{1+z}\over9}\right)^{-1/2}
L_{\alpha,47}^{1/2}.
\end{equation}
For the best case of $T_\alpha\gg T_\CBR$, the efficiency for 21-cm emission
may be expressed, using equations~(\ref{eq:ya}), (\ref{eq:ftspin}),
(\ref{eq:ptherm}), and (\ref{eq:eff}), as $\eta=(1+r^2/r^2_{\rm th})^{-1}$.
At $r=r_{\rm th}$, $\eta=1/2$. Comparison with equation~(\ref{eq:rieq}) shows
that the thermalization distance is comparable to the size of the \HII region.
Since $r>r_I$ is required for detection, $\eta>1/2$ may be achieved only for
a brief redshift interval after the quasar turns on. This interval is so
small for many quasars that their \HII regions will still be in the luminal
expansion stage (cf. equation~[\ref{eq:rcrit}]), so that they will not be
preceded by a warming front and will thus not be visible in 21-cm emission.
The efficiency will therefore generally lie in the range $0<\eta<1/2$.

Larger shells offer a slight advantage for detectability. This is because
the flux of a shell emitting 21-cm radiation depends not only on its
efficiency but on its size as well. The frequency integrated flux is
proportional to
\begin{equation}
\left(\Omega \Delta \nu \right)_{\rm eff}\equiv 2\pi\int d\nu
\int {d\theta \theta \eta}. \label{eq:Oeff}
\end{equation}
Consider a measurement made at a frequency corresponding to a depth in the
emitting zone around the quasar just outside the I-front, and with a bandwidth
$\Delta\nu_c$ matched to the light radius $r_c$. Denoting by $\theta_{\rm th}$
the angle on the sky subtended by $r_{\rm th}$, integrating
equation~(\ref{eq:Oeff}) out to $r_c$, using the expression above for $\eta$,
gives $(\Omega\Delta\nu)_{\rm eff}
\simeq2\pi
\theta_{\rm th}^2\Delta\nu_c[1-(r_{\rm th}/r_c)\arctan(r_c/ r_{\rm th})]$, when
$r_c\gg r_I$. While the signal increases initially like the cube of the size
of the emitting region, it does so only linearly at later times. Since the
noise increases like $r_c^{3/2}$, the signal-to-noise ratio increases
asymptotically only like $r_c^{1/2}$. Allowing for a light radius that subtends
an angle $\theta_c$ on the sky of $1^\circ$, the flux is then
\begin{equation}
\delta S\approx (1-2~{\rm mJy\,}) h_{50}\left({\Omega_{\rm
IGM}h_{50}^2\over 0.05}\right) \left({{1+z}\over 9}\right)^{-1/2}
L_{\alpha, 47}
\end{equation}
over a bandwidth of $\Delta\nu_c\approx11\, {\rm MHz\,} [(1+z)/9]^{-1/2}
(\theta_c/1^\circ)$, comparable to the rms noise. It would thus take a few
hundred hours of integration time on the GMRT to detect the shell at
$5\sigma_{\rm rms}$. A quicker strategy may be to search for an excess variance
across the sky over the rms noise that these shells would produce in the
observed flux per resolution (beamwidth$\times$bandwidth) element. A next
generation facility like the {\it Square Kilometer Array Interferometer}
(Braun 1996) could easily map warm IGM bubbles, effectively opening much of the
universe to a direct study of the reheating epoch.  

\section{Summary}

The spectral appearance at 21-cm of non-uniform gas at high-$z$ had been
previously investigated by Hogan \& Rees (1979) and Scott \& Rees (1990). 
These earlier papers showed that the 21-cm emission  from high-$z$  \HI
may display angular structure, as well as structure in redshift space;  radio
observations could therefore in principle yield ``tomographic'' mapping of
protoclusters or other large-scale inhomogeneities  at high redshifts,  where
the gas had not yet been reionized. The present work discusses how similar
techniques may be used to reveal ``patchiness'' in the 21-cm emission due to
non-uniformities in the spin temperature, rather than in the density. Such
effects could tell us how  (as well as when) the primordial gas was reheated;
if the reheating were due to ultraluminous but sparsely distributed sources
(e.g., the first QSOs), the resultant patches would have a larger scale than
any gravitationally-induced inhomogeneities, and would therefore be less
difficult to detect. The continuum radiation, redshifted to \Lya, from an early
generation of stars would couple the spin temperature to the low temperature of
the IGM for a transitory interval before heating the gas above the CBR
temperature. The result is a ``window of opportunity'' during which it may be
possible to detect the IGM in absorption against the CBR, and in so doing
possibly reveal the first epoch of star formation.

We have assessed three preheating mechanisms at early times: soft X-ray
radiation from quasar sources, soft X-rays from virialized
$10^{10}-10^{11}\msun$ halos, and resonant scattering of background \Lya
continuum photons from an early generation of stars, and argued that radiation
sources at high redshifts may prevent collapsing structures that
cannot cool efficiently from being detected in 21-cm absorption
against the microwave background. We have shown that scattering of \Lya 
continuum radiation can mix the hyperfine levels of intergalactic \HI within
$\sim 10\,$Mpc of an individual QSO, and that this will produce an
appreciable 21-cm emissivity from the IGM in the quasar neighborhood. While the
X-ray-heated bubbles will still survive as fossils, once the quasar has
died, the \Lya photons will rapidly adiabatically redshift out of resonance
so that no emission will occur without an external supply of photons.

If reionization occurred at redshift $5\lsim z\lsim 10$, these QSO spheres of
influence might be detectable at meter wavelengths by the GMRT. Because of the
very large scales involved, there is no difficulty in achieving a sufficiently
narrow bandwidth and an adequate angular resolution with the GMRT; the main
limitations are the low flux levels and the radio background subtraction. 

It is possible that both the reionization and heating of the IGM were due to
an early population of stars on subgalactic scales (rather than quasars) at 
redshift $z\gsim 20$, well beyond the reach of the GMRT. Even if they reionized
the IGM at an epoch accessible to the GMRT, however, these sources would be less
thinly spread, and the characteristic structures in the IGM generated by them
would be smaller and less detectable, so that an alternative observing strategy
than the one discussed here would be required. 

\acknowledgments
We acknowledge helpful conversations on the subject of this paper with 
P. Palmer, L. Spitzer, M. Voit, W. Watson, and M. White. We thank the referee,
G. Swarup, for valuable comments which helped to clarify the detectability of
a neutral IGM with the GMRT. P.M. and A.M. were supported in part by NSF grant
PHY94-07194 to the ITP, University of California, Santa Barbara. A.M. is also
grateful to the W. Gaertner fund for its support. M.R. thanks the Royal Society
for support. 

\appendix

\section{\Lya Line Flux in the Neutral Medium}

We evaluate the \Lya flux produced in the neutral \HI outside an ionization
front from recombinations within the \HII region itself. 
The volume emissivity is given by 
$$
\epsilon_\nu=x_e x_p n_\nH^2 \alpha^{\rm eff}_{2^2P}h\nu\phi(\nu), \eqno({\rm 
A1})
$$
where $\alpha^{\rm eff}_{2^2P}$ is the effective recombination coefficient 
to the $2^2P$ level (Pengelly 1964). Integrating from just outside the
ionization front into the \HII region, and averaging over $4\pi$ sr, we get 
$$
J_\alpha(r)={1\over 4\pi}\int_{r_I}^r{\epsilon_\nu dr}
\approx 1.2\,\alpha_{2^2P}^{\rm eff} {n_\nH^2 hc\over 8\pi H} \eqno({\rm A2})
$$
(with $x_e=1.2$, $x_p=1$, and including an additional factor of $1/2$ since 
\Lya photons arise only from within the \HII region). In evaluating the
integral, we relate the distance $r-r_I$ from the ionization front to the
frequency shift $y$ from line center (in units of the thermal width) due
to the Hubble expansion, by $r-r_I=by/H$, where $b$ is the Doppler parameter.
Because emission at line center is quickly redshifted out of the core, only the
surface of the \HII region contributes to the \Lya flux. The resulting
intensity can be expressed as
$$
J_{\alpha,-21}(z)\approx 0.03\, h_{50}^{-1}
\left({\Omega_{\rm IGM}h_{50}^2\over 0.05}\right)^2 
\left({{1+z}\over 7}\right)^{4.5}. \eqno({\rm A3})
$$
The recombination photons escaping from an expanding \HII region produce 
then only a small flux of \Lya photons. 

For completeness, we shall also compute the contribution from the local
production of \Lya photons in the warm (mostly) neutral material surrounding
the ionized gas. As \Lya photons are redshifted out of the Doppler core by the
Hubble expansion, the local background intensity becomes
$J_\alpha=$ $(n_\nH^2hc/4\pi H)$ $[x_ex_p\alpha^{\rm
eff}_{2^2P}+$ $x_ex_\nHI\gamma_{\rm eH}+$ ${\cal E}{\dot E_X}/(n_\nH
h\nu_{\alpha})]$, where $\gamma_{\rm eH} \approx (2.2\times10^{-8}\,{\rm cm^3
s^{-1}})\exp(-11.84/T_4)$ (Osterbrock 1989) is the collisional excitation rate
of \Lya photons by electron impacts, and the last term accounts for \Lya
collisional excitations from the nonthermal distribution of electrons generated
by the soft X-rays. The factor ${\cal E}\approx 0.4$ (Shull \& van
Steenberg 1985) represents the net fraction of photon energy converted to
\Lya photons. For a temperature of $T_4=0.1$ and allowing (optimistically) for
$x_e\sim x_p\sim0.2$, radiative recombinations dominate, and 
$$
J_{\alpha,-21}(z)\approx 0.5\, x_e x_p h_{50}^{-1}
\left({\Omega_{\rm IGM}h_{50}^2\over 0.05}\right)^2 
\left({{1+z}\over 7}\right)^{4.5}. \eqno({\rm A4})
$$
Recombinations and collisional excitations in a warm, largely neutral medium
are therefore a negligible source of \Lya photons. 

\section{Heating by Resonant Scattering of \Lya Photons}

We derive the rate of transfer of energy, equation~(\ref{eq:edot}), from the 
radiation field to the atoms through recoil using the equation of  
radiative transfer. Neglecting expansion and source terms, the rate of
change of the number density of photons in a given frequency band $n_\nu$ for
an isotropic radiation field was shown by Basko (1981) to be given in the 
diffusion approximation by
$$
{1\over c}{{\partial n_\nu}\over{\partial t}}={1\over2}n_\nHI\Delta\nu_D
{\partial\over{\partial\nu}}\left[\sigma_\nu\left(\Delta\nu_D
{{\partial n_\nu}\over{\partial\nu}} + 2fn_\nu\right)\right],
\eqno({\rm B1})
$$
where $n_\nHI$ is the number density of hydrogen atoms in the ground state,
$f=h\nu_{\alpha}/m_\nH cb$ expresses the effect of recoil,
and only terms of order $v/c$ have been retained. This is only
correct as long as $kT_K\ll h\nu_{\alpha}$. Note that, in a steady state,
$n_\nu\propto\exp[-h(\nu-\nu_{\alpha})/kT_K]$ near the resonant frequency, 
as was first demonstrated by Field (1959b).

The rate of change of energy in the radiation field due to recoil is then
$$
\dot u_\alpha = h\nu_\alpha \dot n_\alpha = n_\nHI c\Delta\nu_Dh\int_0^\infty
{d\nu\ \nu {\partial\over{\partial\nu}}\left(\sigma_\nu fn_\nu\right)}.
\eqno({\rm B2})
$$
Performing the integration by parts, noting that $\sigma_\nu$ is sharply
peaked at the resonant \Lya frequency, using $n_\nu=(4\pi/c)J_\nu/h\nu$, and
setting $\dot E_\alpha=-\dot u_\alpha/n_\nHI$, we recover
equation~(\ref{eq:edot}) of the text. This result is valid in the presence
of expansion and source terms as
well, as may be shown from the generalization of equation (B1) by Rybicki \&
Dell'Antonio (1994). We note that, to order $v/c$, the flow of energy is
one-directional, from the radiation field to the atoms. Treating the reverse
flow requires the inclusion of terms of order $(v/c)^2$ (including relativistic
effects like aberration). We have also neglected stimulated emission. These
effects become particularly important when the matter approaches thermal
equilibrium with the radiation field. In this limit, the radiation field
relaxes to a blackbody spectrum, and the net transfer of energy ceases. In the
situations relevant to this paper, these effects are negligible. 

\references

Adams, T.~F. 1971, \apj, 168, 575

Allison,, A.~C., \& Dalgarno, A. 1969, \apj, 158, 423

Allison, A.~C., \& Smith, F.~J. 1971, Atomic Data, 3, 317

Arons, J., \& Wingert, D.~W. 1972, \apj, 177, 1

Bahcall, J.~N., \& Ekers, R.~D. 1969, \apj, 157, 1055 

Bardeen, J.~M., Bond, J.~R., Kaiser, N., \& Szalay, A.~S. 1986,
\apj, 304, 15

Basko, M.~M. 1981, Astrophysics, 17, 69

Bebbington, D.~H.~O. 1986, \mnras, 218, 577

Bennett, C.~L., \etal 1994, \apj, 434, 587

Bergeron, J., \& Souffrin, S. 1971, \aap, 11, 40

Bond, J.~R. 1995, in Cosmology and Large Scale Structure, ed. R.
Schaeffer (Netherlands: Elsevier Science Publishers), in press

Braun, R. 1996, in Cold Gas at High Redshift, eds. M. Bremer
\etal (Kluwer: Dordrecht), in press

Couchman, H.~M.~P. 1985, \mnras, 214, 137

Couchman, H.~M.~P., \& Rees, M.~J. 1986, \mnras, 221, 53

Deguchi, S., \& Watson, W.~D. 1985, \apj, 290, 578

Fabbiano, G. 1989, \araa, 27, 87

Field, G.~B. 1958, Proc. I.R.E., 46, 240

--------- . 1959a, \apj, 129, 536

--------- . 1959b, \apj, 129, 551

%Fukugita, M., \& Kawasaki, M. 1994, \mnras, 269, 563

Gallego, J., Zamorano, J., Arag\' on-Salamanca, A., \&
Rego, M. 1995, \apjl, 455, L1

G\' orski, K.~M., Ratra, B., Sugiyama, N., \& Banday, A.~J.
1995, \apjl, 444, L65 

Gould, R.~J., \& Thakur, R.~K. 1970, Ann. of Phys., 61, 351

Gunn, J.~E., \& Peterson, B.~A. 1965, \apj, 142, 1633

Hogan, C.~J., \& Rees, M.~J. 1979, \mnras, 188, 791

Hu, W., \& White, M. 1996, \aap, in press

%Liddle, A.~R., \& Lyth, D.~H. 1995, 273, 1177

Longair, M.~S. 1995, in Extragalactic Background
Radiation, eds. D. Calzetti, M. Livio, \& P. Madau (Cambridge: Cambridge
University Press), p. 223

Madau, P., \& Meiksin, A. 1991, \apj, 374, 6

--------- . 1994, \apjl, 433, L53

Madau, P., \& Shull, J.~M. 1996, \apj, 457, 551

Mather, J.~C., \etal 1994, \apj, 420, 439

Meiksin, A. 1994, \apj, 431, 109

Meiksin, A., \& Madau, P. 1993, \apj, 412, 34

Miralda-Escud\'e, J., \& Rees, M.~J. 1994, \mnras, 266, 343

Moiseiwitsch, B.~L. 1962, in Atomic and Molecular Processes,
ed. D.~R. Bates\ (New York: Academic Press), p. 280

Osterbrock, D.~E. 1989, Astrophysics  of Gaseous Nebulae and
Active Galactic Nuclei (Mill Valley: University Science Books)

Peebles, P.~J.~E. 1993, Principles of Physical Cosmology,
(Princeton: Princeton University Press), Ch. 6

Pengelly, R.~M. 1964, \mnras, 127, 145

Phinney, E.~S. 1985, Astrophysics of Active Galaxies and QSOs,
ed. J. S. Miller (Oxford: Oxford University Press), p. 453

Press, W.~H., \& Schechter, P. 1974, \apj, 187, 425

Purcell, E.~M., \& Field, G.~B. 1956, \apj, 124, 542

Rees, M.~J. 1985, \mnras, 213, 75P

Rybicki, G.~B., \& Dell' Antonio, I.~P. 1994, \apj, 427, 603

Schneider, D.~P., Schmidt, M., \& Gunn, J.~E. 1991, \aj, 101, 2004

Scott, D., \& Rees, M.~J. 1990, \mnras, 247, 510

Shapiro, P.~R. 1986, \pasp, 98, 1014

Shull, J.~M., \& van Steenberg, M.~E. 1985, \apj, 298, 268

Spitzer, L. 1962, Diffuse Matter in Space,
(New York: Wiley-Interscience)

Spitzer, L. 1978, Physical Processes in the Interstellar Medium,
(New York: Wiley-Interscience)

Spitzer, L., \& Scott, E.~H. 1969, \apj, 158, 161

Subramanian, K., \& Padmanabhan, T. 1993, \mnras, 265, 101

Sugiyama, N., Silk, J., \& Vittorio, N. 1993, \apjl, 419, L1

Swarup, G. 1996, in Cold Gas at High Redshift, eds. M. Bremer
\etal (Kluwer: Dordrecht), in press

Swarup, G., Ananthakrishan, S., Kapahi, V.~K., Rao, A.~P.,
Subrahmanya, C.~R., \& Kulkarni, V.~K. 1991, Curr. Sci., 60, 95

Swarup, G., \& Malik, R.~K. 1996, private communication

%Tegmark, M., Silk, J., \& Blanchard, A. 1994, \apj, 420, 484

Urbaniak, J.~J., \& Wolfe, A.~M. 1981, \apj, 244, 406

Warren, S.~J., Hewett, P.~C., \& Osmer, P.~S. 1994, \apj, 421, 412

White, S.~D.~M., Efstathiou, G., \& Frenk, C.~S. 1993, \mnras,
262, 1023

Wouthuysen, S.~A. 1952, \aj, 57, 31

\newpage

\figcaption{ Numerical calculation of the IGM surrounding a QSO
(assumes $q_0=0.5$, $H_0=50\kmsmpc$, $\Omega_{\rm IGM}=0.05$). The QSO was
turned on gradually at $z=6$ with a specific luminosity of
$L_\nu=10^{31}\,$ ergs $s^{-1}$ $Hz^{-1}$ $(\nu/\nu_L)^{-1.5}$. The effects of
secondary electrons are significant for soft X-rays, and have been included,
along with radiative cooling mechanisms and Compton cooling off the CBR.\ (a)
Shows the temperature of the IGM due to soft X-ray heating by the QSO. At 
$z=5.5$, the IGM is heated to a temperature above that of the cosmic 
background radiation out to a radius of 23 Mpc. The light front is at 25 Mpc.\ 
(b) Shows the \HII ({\it solid}) and
\HeII ({\it dashed}) fractions. The ionized region surrounding the QSO extends
to 7.5 Mpc at this time. \label{fig1}}

\figcaption{ Heating rate per hydrogen atom of the IGM due to soft X-rays from
collapsed galaxy halos, in a flat $q_0=0.5$ and $H_0=50\kmsmpc$ CDM-dominated
cosmology ({\it solid}). Also shown is the required heat input over
a Hubble time to match the CBR temperature, $1.5 T_\CBR(z)/ H(z)$
({\it dotted}). For $z<7$, the rate of X-ray heating from collapsed halos is
sufficient to raise the temperature of the IGM above that of the CBR. At
higher redshifts, the IGM is too dense and the hot halos that can produce
the required penetrating soft X-rays are too few for efficient heating.
At these epochs, only part of the IGM will be heated above the CBR temperature.
The heating rate $\dot E_{\rm th}$ due to \Lya resonant scattering ({\it
dashed}) is shown for 
a background radiation field sufficiently strong to couple the spin temperature
to the kinetic temperature of the IGM (see text). \label{fig2}} 

\figcaption{ (a)\ The 21-cm efficiency is shown as a function of distance
from a QSO with $L_{\alpha, 47}=1$ and $\alpha_S=1.5$, turning on at $z=8.7$.
A \Lya radiation field of intensity $J_{\alpha, -21}=0.8[(1+z)/9]$ was turned
on at $z=9$ and partially couples the spin temperature of the neutral hydrogen
to the kinetic temperature. The effect of heating of the IGM by \Lya collisions
is included. The efficiency is shown at time intervals of 2.5, 5, and 10$\times
10^7$~yr after QSO turn-on, corresponding to $z=8.3$, 8.0, and 7.4,
respectively. Because the IGM temperature is less than that of the cosmic
background radiation at early times, a large efficiency is achieved in an X-ray
warmed shell surrounding the quasar.\ (b)\ A map of the differential antenna
temperature (K) as a function of angle (degrees) and velocity shift ($10^3
\kms$) (differenced from 21-cm redshifted to 153 MHz), resulting from two
quasars turning on at $z=8.7$, as in (a). One quasar is at a comoving position
corresponding to $z=7.4$ and the other at $z=8.0$. The quasars are separated by
$2.5^\circ$ on the sky. Such a map could be constructed by making a sequence of
angular maps in a given frequency band, here taken to be $2000\kms$ wide, and
stepping in frequency until the full emitting region surrounding the quasars is
recovered. In practice, a bandwidth of $5000-10000\kms$ will give a more
feasible integration time for establishing a positive detection. The emission
is measured relative to a reference beam well-separated from the quasars,
either by frequency or by angle. \label{fig3}} 

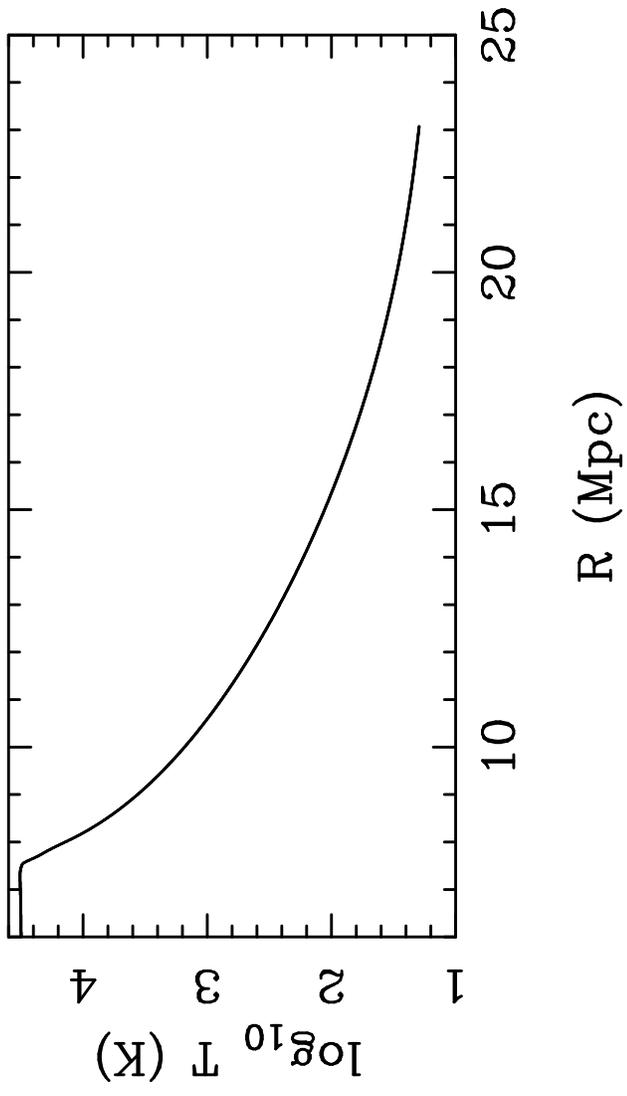
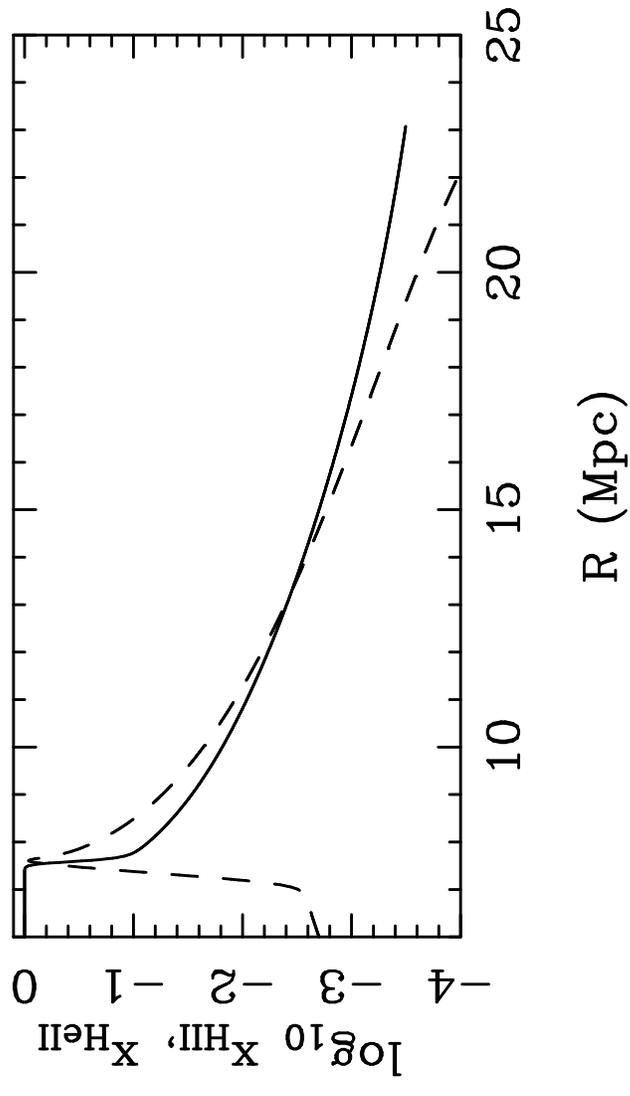
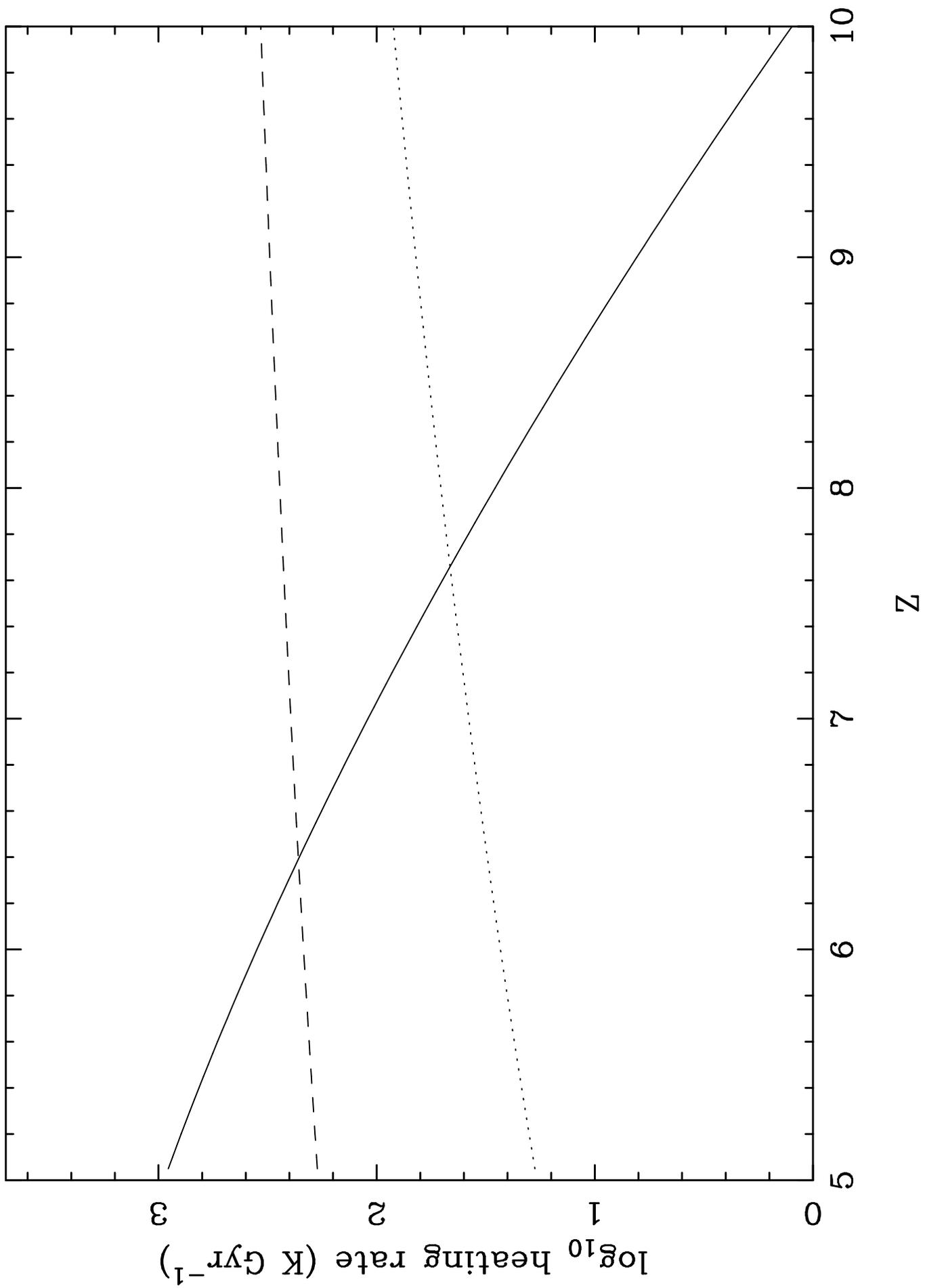
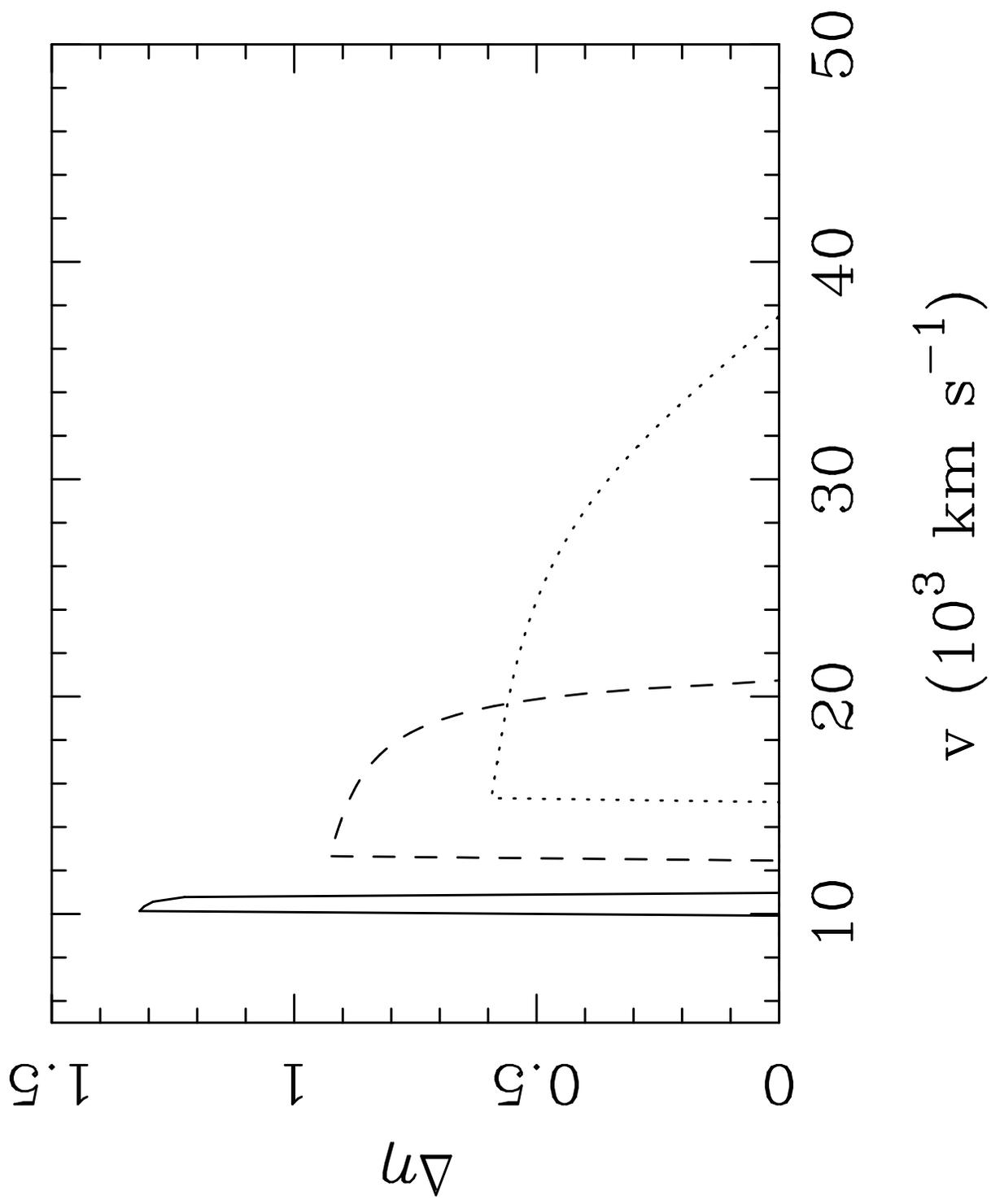
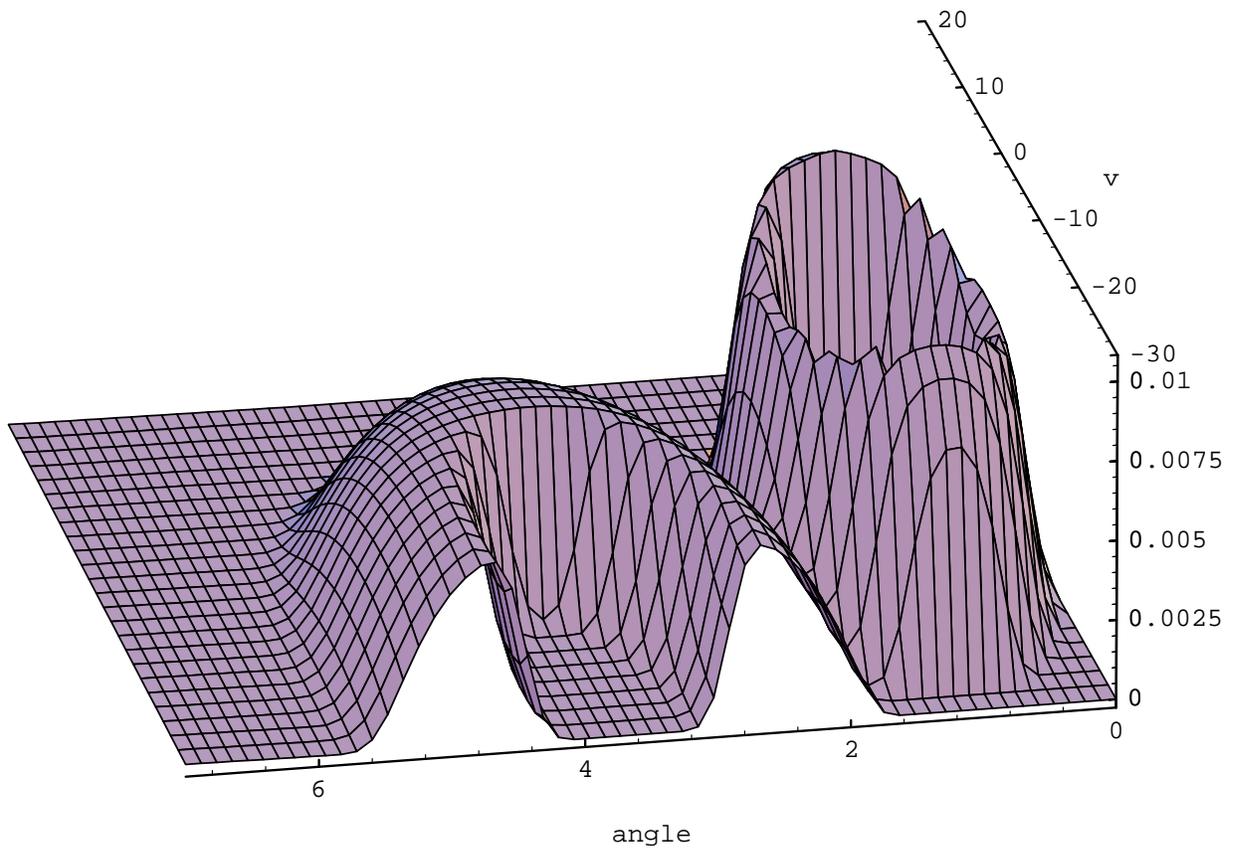
\end{document}